\documentclass[epjCONF]{svjour}
\usepackage{graphics}
\usepackage{graphicx}
\usepackage[varg]{txfonts} 
\usepackage[latin1]{inputenc}
\usepackage{lineno}
\session-title{UHECR2012}
\begin{document}
\title{Review of the Anisotropy Working Group at UHECR-2012}
\author{
O. Deligny\inst{1}, J. de Mello Neto\inst{2} \and P. Sommers\inst{3}, for the Pierre Auger Collaboration\\
H. Sagawa\inst{4}, P. Tinyakov\inst{5,6} \and I. Tkachev\inst{5}, for the Telescope Array Collaboration\\
A. Ivanov\inst{7} \and L. Timofeev\inst{7} for the Yakutsk Array Group
}
\institute{IPN Orsay, CNRS/IN2P3 \& Universit\'e Paris 11, Orsay, France \and Universidade Federal do Rio de Janeiro, Instituto de F\'isica, Rio de Janeiro, RJ, Brazil \and Pennsylvania State University, University Park, PA, USA \and Institute for Cosmic Ray Research, University of Tokyo, Kashiwa, Chiba, Japan \and Institute for Nuclear Research of the Russian Academy of Sciences, Moscow, Russia \and Universit\'e Libre de Bruxelles, Brussels, Belgium \and Shafer Institute for Cosmophysical Research \& Aeronomy, Yakutsk, Russia}
\abstract{
The study of ultra-high energy cosmic rays (UHECRs) has recently experienced a jump in statistics 
as well as improved instrumentation. This has allowed a better sensitivity in searching for anisotropies
in the arrival directions of cosmic rays. In this written version of the presentation given by the 
inter-collaborative "Anisotropy Working Group" at the International Symposium on Future
Directions in UHECR physics at CERN in February 2012, we report on the current status for
anisotropy searches in the arrival directions of UHECRs. 
} 
\maketitle


\section{Introduction}
\label{intro}

At the International Symposium on Future Directions in UHECR physics at CERN in February 2012 (UHECR 2012),
possible future directions of the field of UHECRs were discussed, bringing the major collaborations from air-shower
experiments as well as colleagues from theory together. An inter-collaborative "Anisotropy Working Group" was
formed, with the task to compile a balanced view about the current situation, to identify critical points, and to
give some perspectives about future challenges. 

UHECR anisotropy is a broad topic. Around $10^{18}~$eV, interesting constraints come from measuring the 
first harmonic modulation in the right ascension distribution of cosmic ray arrival directions. The search
for point-like sources that would be indicative of a flux of neutrons in this energy range is also interesting.
At higher energies, searches for clustering in arrival directions, as well as searches for correlations between 
cosmic ray arrival directions and nearby extragalactic objects or the large scale structure of the Universe
are the best suited tools in attempting to understand further the origin of UHECRs.

We describe in the following the current status of each of these sub-topics in the UHECR anisotropy. 

\section{Autocorrelations and harmonic analyses}
\label{sec:2}
\subsection{Autocorrelations}
\label{subsec:21}
The deflections imprinted by the intervening galactic and extragalactic magnetic fields in the arrival directions of
cosmic rays prevent them from pointing back to the sources. The intensity and orientation of these fields are not well
known, but as the deflections decrease with the inverse of the rigidity, the effect is smaller at the highest energies. 
Thus, it is at the highest energies that cosmic rays are most likely to point towards their sources. Meanwhile, if the 
suppression of the cosmic ray intensity at the highest energies is due to the interactions of UHECRs with the cosmic
microwave background, cosmic rays with energies above $\simeq$ 60~EeV are expected to come mostly from nearby 
sources. These ideas have motivated an extensive search for clustering signals both at small angular scales, looking 
for point-like sources, and at intermediate angular scales, looking for the pattern characterizing the distribution of nearby
sources. 

A standard tool for studying clustering is provided by the autocorrelation function. This counts the number 
of pairs separated by less than an angle $\delta$ among the events with energy larger than an energy threshold $E$. 
The expected number of pairs can be obtained by generating a large number of Monte Carlo simulations with the same 
number of events as in the data set with an isotropic distribution modulated by the exposure of the detector, from which the mean number 
of pairs and dispersion regions can be extracted for each angular scale. The fraction of simulations with a larger 
number of pairs than the data gives a measure of the probability that an observed excess of pairs arises by chance from 
an isotropic distribution of events. Small scale clustering at energies larger than 40~EeV has been claimed by the AGASA 
experiment~\cite{AGASA}. However, this claim has not been supported by blind tests on independent data sets provided 
by other experiments. The HiRes experiment has found no significant clustering signal at any angular scale up to 5$^\circ$ 
for any energy above 10~EeV~\cite{HiRes}. Below, we report in more details on autocorrelation analyses performed 
on TA and Auger data.

\begin{figure}
\center{
\resizebox{0.32\columnwidth}{!}{  \includegraphics{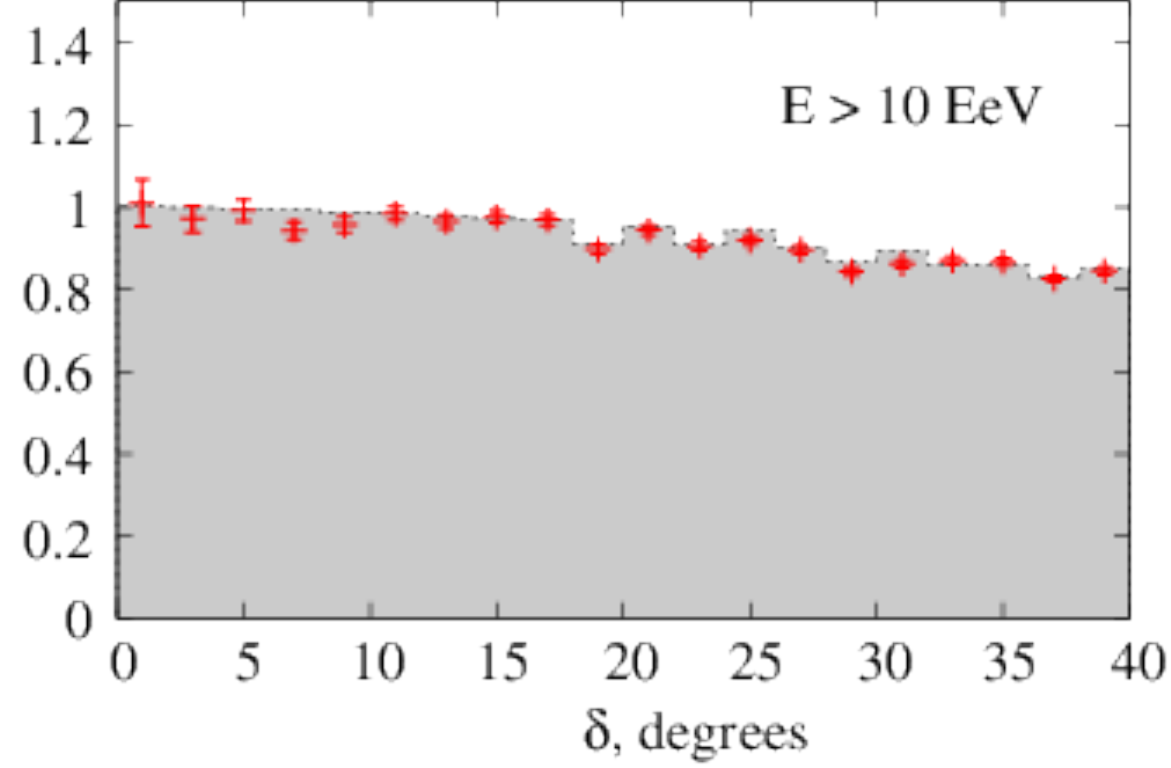} }
\resizebox{0.32\columnwidth}{!}{  \includegraphics{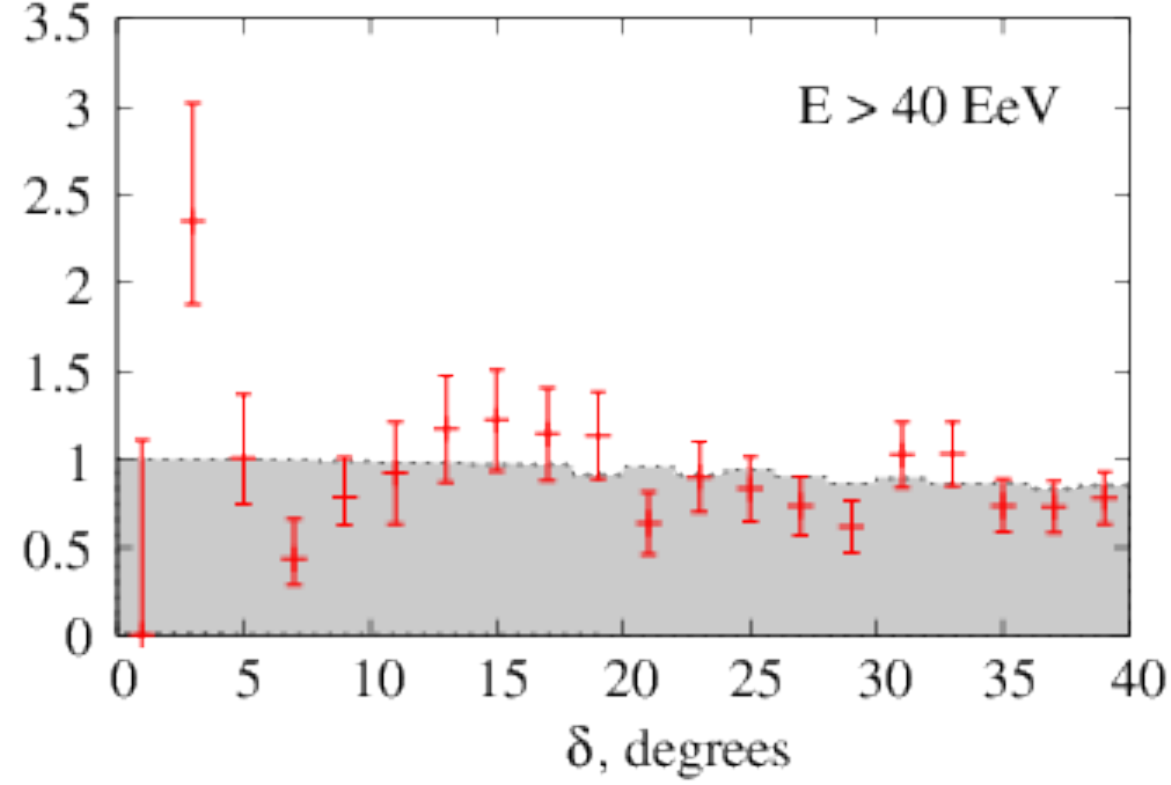} }
\resizebox{0.32\columnwidth}{!}{  \includegraphics{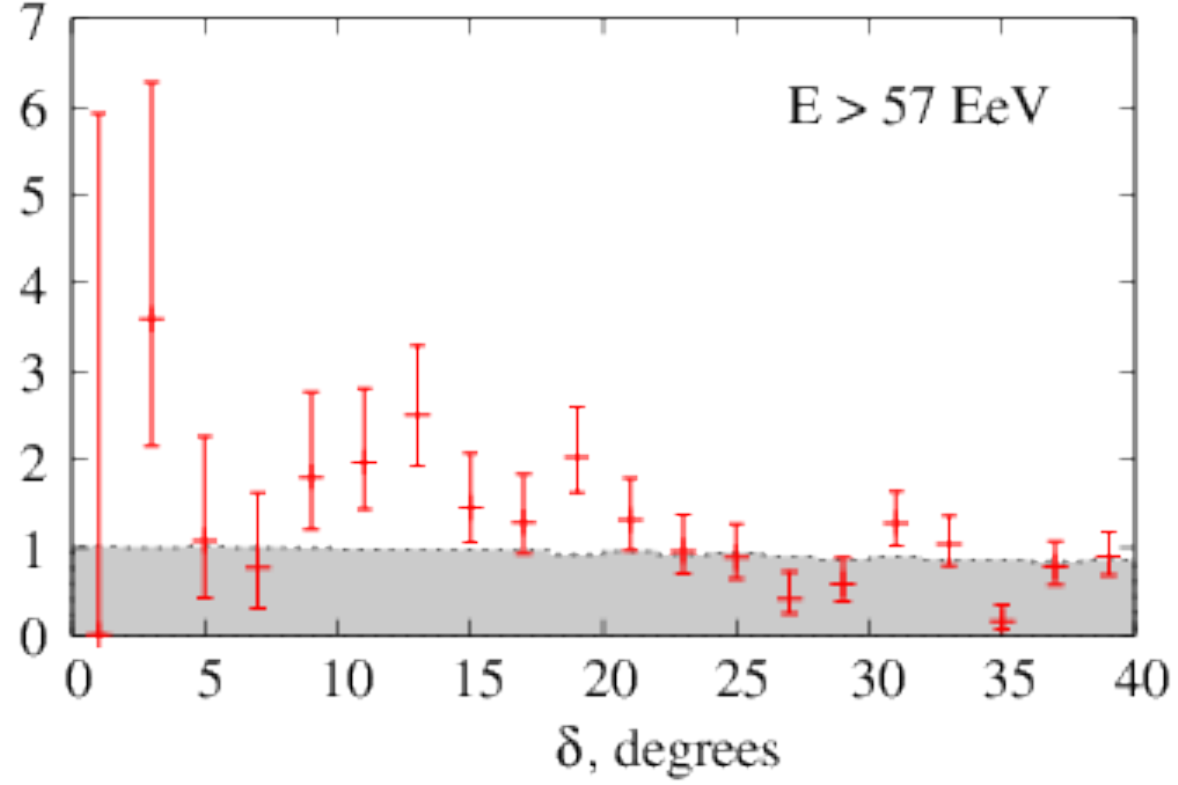} }
}
\caption{Number of pairs in the TA data set with angular separation $\delta$ normalised to the area of the angular bin
(data points), compared to the expectation for the uniform distribution (shaded histograms). From left to right~:
$E>10$~EeV, $E>40$~EeV, and $E>57$~EeV. From~\cite{TA}.}
\label{fig:autoTA}
\end{figure}

\begin{figure}\centering
\includegraphics[angle=-90,width=0.45\linewidth]{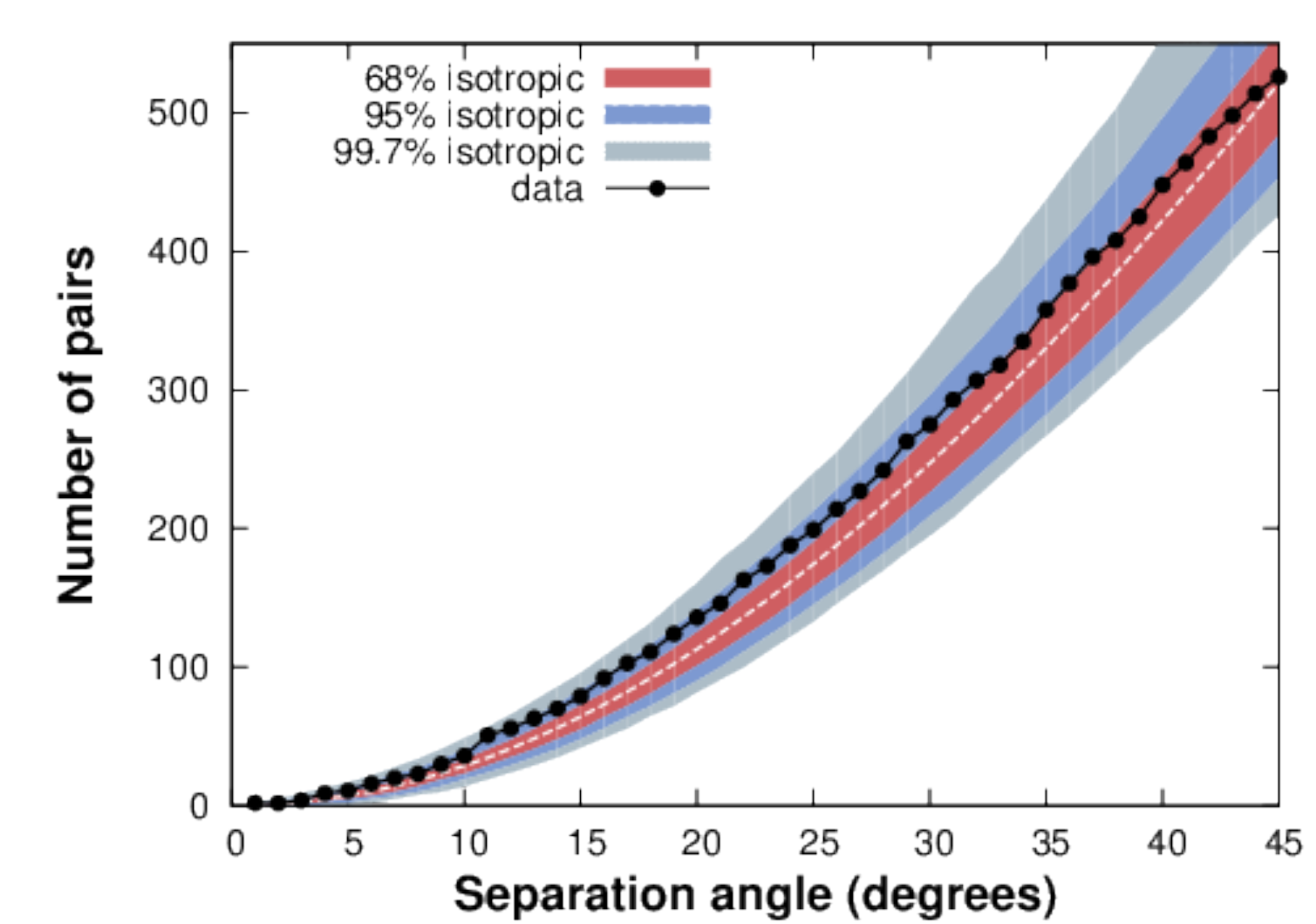}
\includegraphics[angle=-90,width=0.45\linewidth]{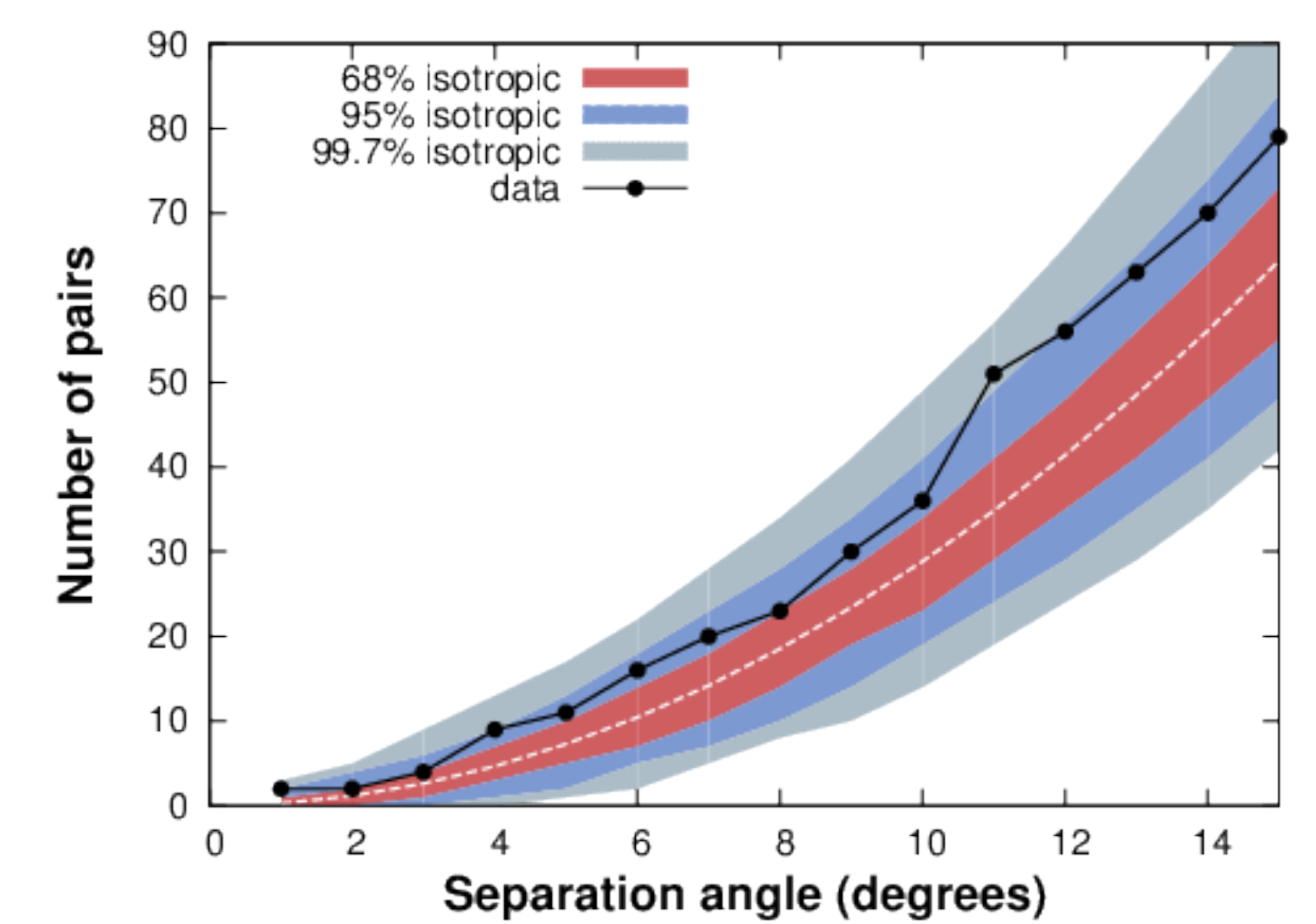}
\caption{Number of pairs in the Auger data set as a function of the angular separation for $E \ge 55$~EeV (black dots)~\cite{Auger1}. 
The bands correspond to the 68\%, 95\% and 99.7\% dispersion expected for an isotropic flux. The plot in the right panel is an enlarged 
version of the left plot restricted to separation angles less than $15^\circ$. }
\label{fig:autoAuger}
\end{figure}

The autocorrelation analysis extended to all angular scales is illustrated in Figure~\ref{fig:autoTA}, using
three energy thresholds of $10$~EeV, $40$~EeV, and $57$~EeV with the TA data~\cite{TA}. 
For each energy threshold, the plot shows the number of pairs with the angular 
separations $\delta$ binned in $2^\circ$ bins (data points). The shaded region represents the average number 
of pairs expected in the case of the uniform distribution. Both the data and the uniform expectation are 
normalized bin-by-bin to the area of the bin, so that in the case of a uniform full-sky exposure the expectation 
would be flat. The overall normalization is set in such a way that the expectation in the first bin equals one.
Throughout the scans in separation angle and in energy,  no significant excess is found. 

The autocorrelation function for the set of 69 events with $E \ge 55$~EeV detected at the Auger Observatory 
is shown in Figure~\ref{fig:autoAuger}~\cite{Auger1}. 
The number of pairs of events with an angular separation smaller than a given value is plotted as black dots. 
The 68\%, 95\%, and 99.7\% confidence intervals expected in the case of an isotropic flux are represented by coloured bands. 
For angles greater than $45^\circ$ (not shown) the black dots lie within the 68\%  band.  The region of small angular 
scale is shown separately for better resolution. The largest deviation from the isotropic expectation occurs for an angular
scale of $11^\circ$, where 51 pairs have a smaller separation compared with 34.8 pairs expected.  In isotropic realizations 
of 69 events, a  fraction $f(11^\circ)=0.013$ have 51 or more pairs within $11^\circ$. Since this angular scale was not
fixed prior to the analysis, there is a statistical penalty factor for choosing this scale \textit{a posteriori}. The fraction of 
isotropic realizations that achieve $f(\psi)\le 0.013$ for any angle $\psi$ is $P=0.10$. There is thus no strong evidence
for clustering at UHE from this analysis. 

Interestingly, although close clusters in both TA and Auger data sets at UHE are absent, one of the TA events is located 
within $1.7^\circ$ from an event detected at the Auger Observatory. Both events have $E>10^{20}$~eV. The center
of the doublet has the Galactic coordinates $l=36^\circ$, $b=-4.3^\circ$.

\subsection{Harmonic analyses}
\label{subsec:22}

Establishing at which energy the flux of extragalactic cosmic rays starts to dominate the cosmic ray energy spectrum 
would constitute an important step forward to provide further understanding on the origin of UHECRs. A time honored 
picture is that the ankle is the onset in the energy spectrum marking the transition between galactic and extragalactic 
UHECRs. As a natural signature of the escape of cosmic rays from the galaxy, diffusion and drift motions could imprint 
dipolar anisotropies in the distribution of arrival directions at the level of a few percent in the energy range just below 
the ankle. Alternatively, if UHECRs above $10^{18}$~eV have already a predominant extragalactic origin, their flux is 
expected to be isotropic to a high level. 

\begin{figure}\centering
\includegraphics[width=0.38\linewidth]{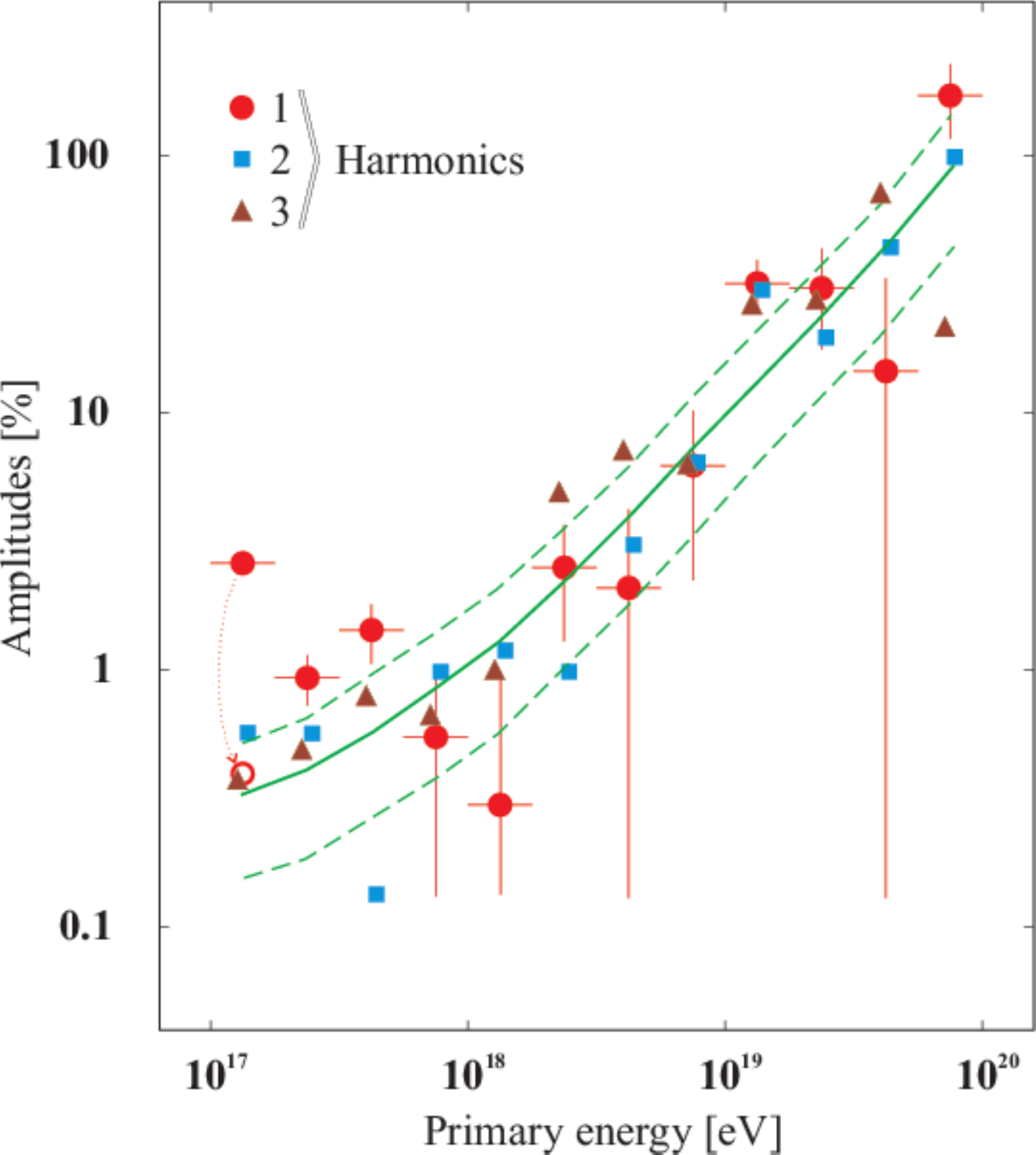}
\includegraphics[width=0.57\linewidth]{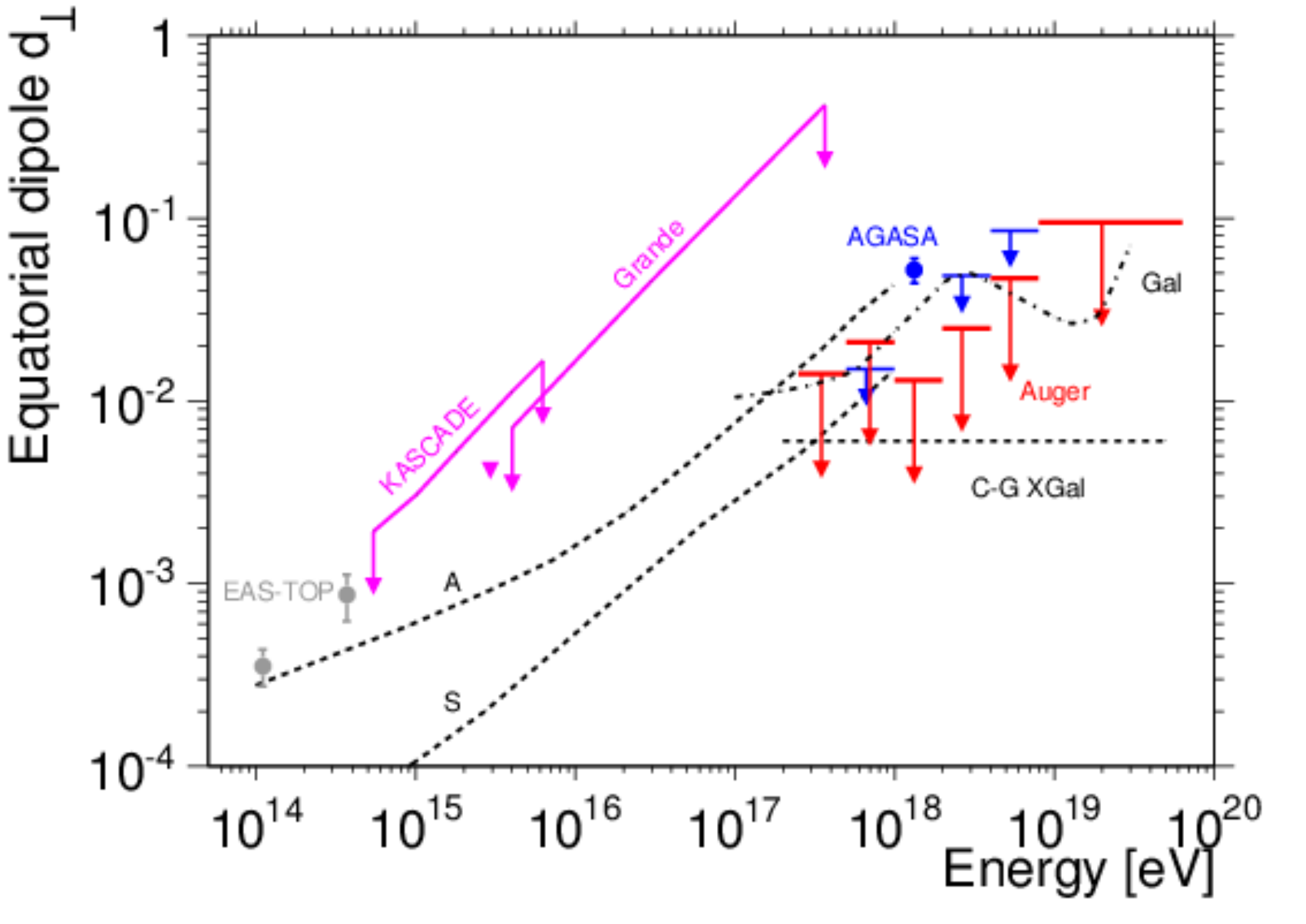}
\caption{Left~: Amplitude of the 1$^{st}$, 2$^{nd}$, and 3$^{rd}$ harmonics as a function of energy for data from 
Yakutsk~\cite{Yakutsk}. Right~: Upper limits on the equatorial dipole component as a function of energy,
from several experiments~\cite{Auger2}. Also shown are the predictions up to 1~EeV from two different galactic magnetic 
field models with different symmetries ($A$ and $S$)~\cite{Roulet}, the predictions for a purely galactic origin of UHECRs 
up to a few tens of $10^{19}\,$eV ($Gal$)~\cite{Calvez}, and the expectations from the Compton-Getting effect for an 
extragalactic component isotropic in the CMB rest frame ($C$-$G\,Xgal$)~\cite{Kachelriess}.}
\label{fig:harmonic1}
\end{figure}

Scrutiny of the large scale distribution of arrival directions of UHECRs as a function of the energy is thus one important 
observable to provide key elements for understanding their origin in the $10^{18}$~eV energy range. Harmonic analysis 
in right ascension benefits from the almost uniform exposure in right ascension of any observatory operating with full
duty-cycle due to the Earth rotation, and constitutes a powerful tool for picking up any dipolar modulation in this coordinate.

\begin{figure}\centering
\includegraphics[width=0.49\linewidth, height=0.33\linewidth]{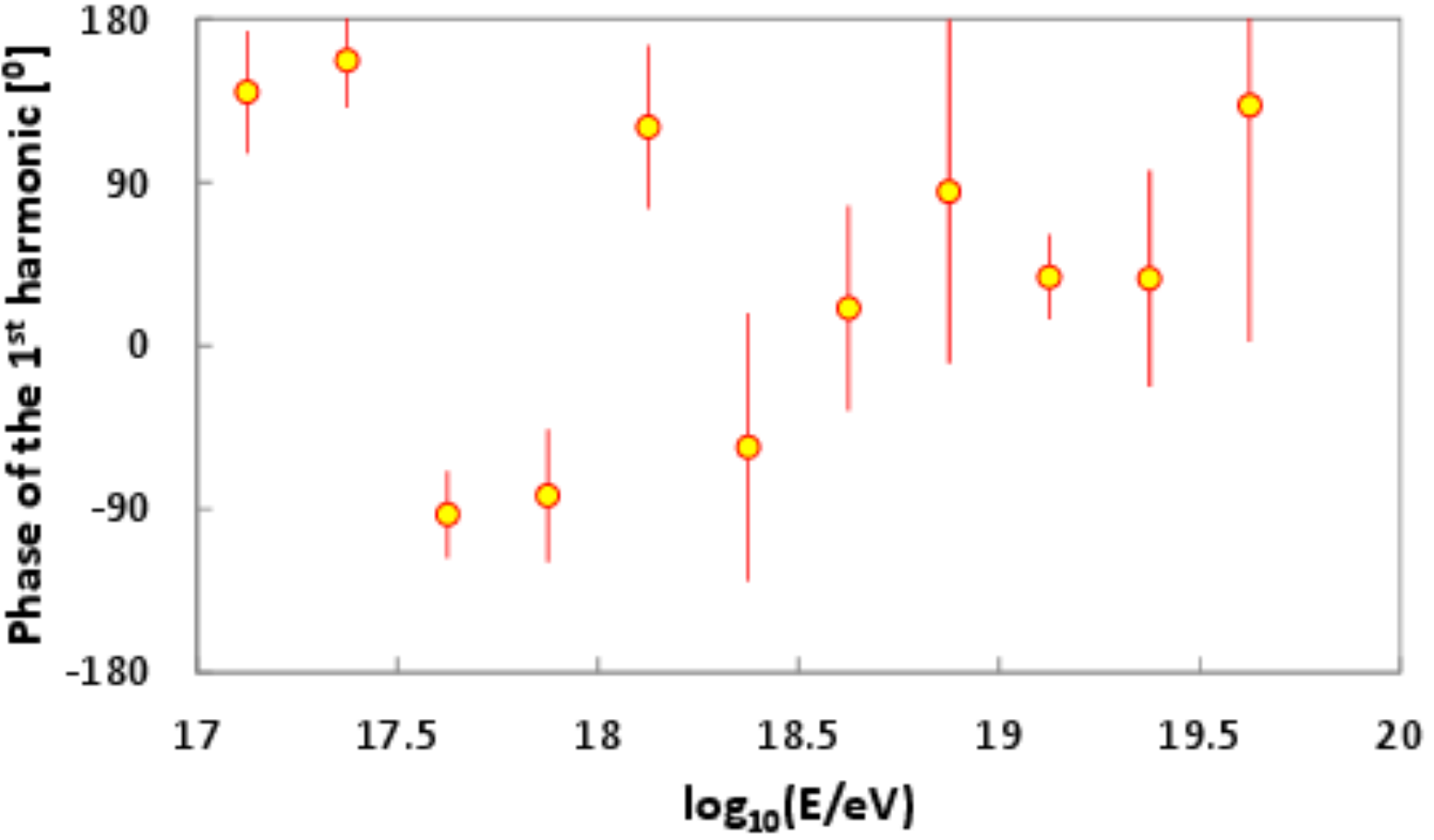}
\includegraphics[width=0.49\linewidth]{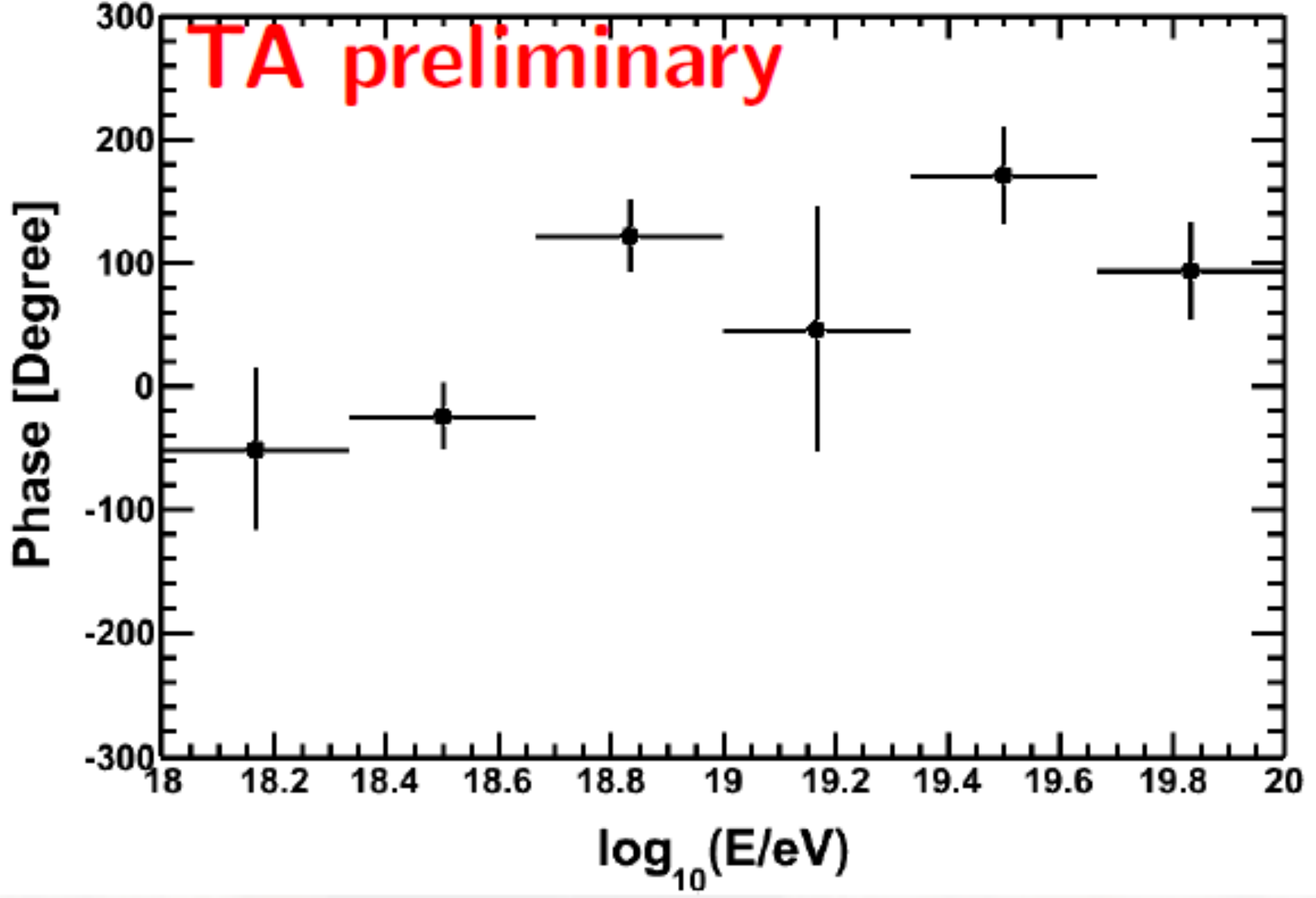}
\includegraphics[width=0.49\linewidth]{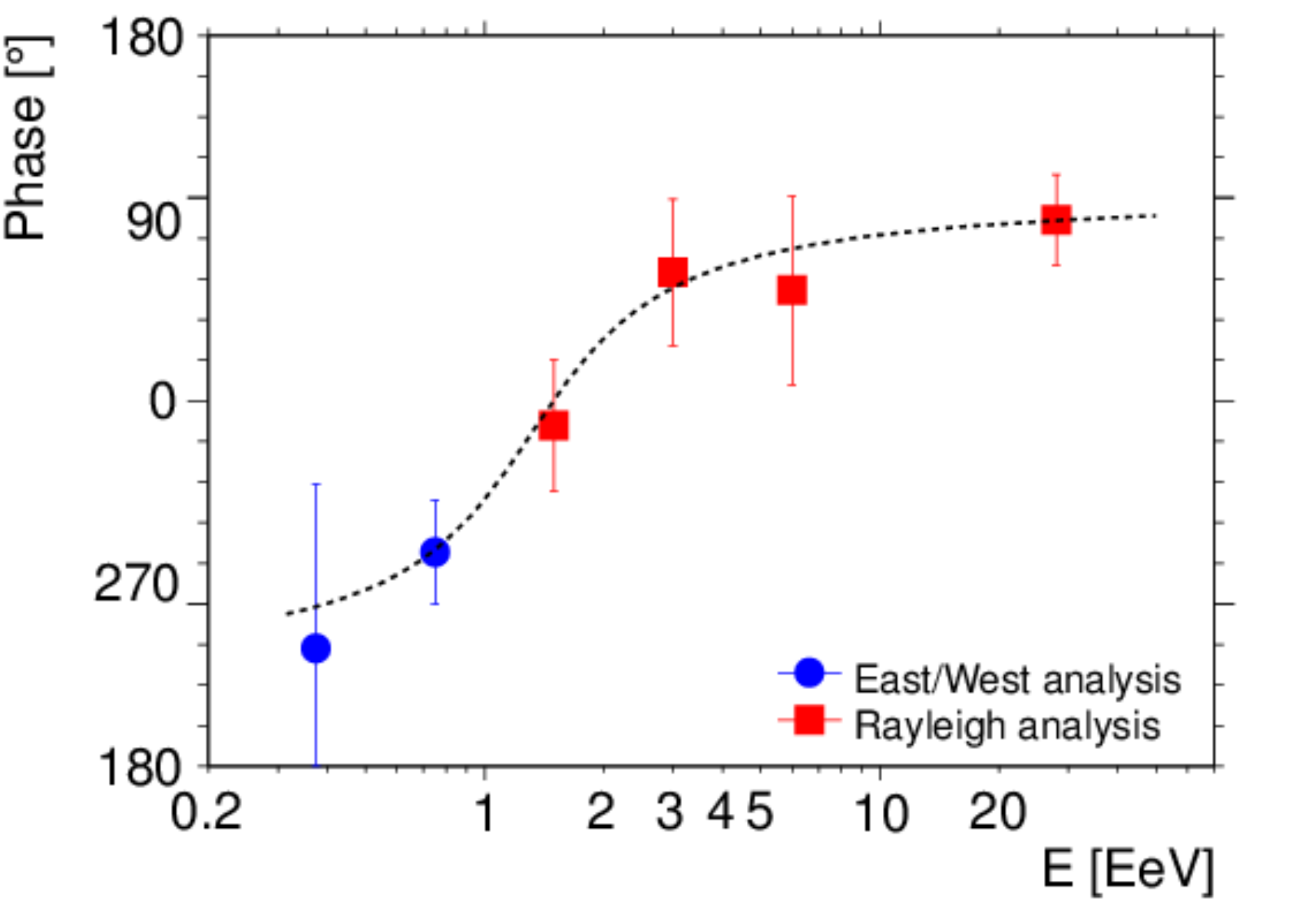}
\caption{Phase of the first harmonics as a function of energy. Top left~: Yakutsk data. Top right~: TA data. Bottom~: Auger data.}
\label{fig:harmonic2}
\end{figure}

The amplitudes of the 1$^{st}$, 2$^{nd}$, and 3$^{rd}$ harmonics as a function of energy for data from Yakutsk~\cite{Yakutsk}
are reported in Figure~\ref{fig:harmonic1} (left panel). One important challenge inherent to this kind of analysis is keeping 
under control the observed rate of events to avoid the false claim of anisotropy. This is particulary important for the first harmonic
analysis, and illustrated with the very left point which shows the raw amplitude observed prior to corrections for seasonal and diurnal
variations of the exposure of the Yakutsk array. The amplitudes and RMS expected for isotropy are shown by the solid and dashed
curves. It is apparent that there is no significant amplitude standing out from the background noise. Similar conclusions were given
in~\cite{Auger2} in this energy range. The corresponding upper limits on the dipole component in the equatorial plane (shown
in the right panel) were obtained, being at the percent level at 99\% C.L. around $10^{18}$~eV and providing the most stringent 
bounds ever obtained.

Besides, it was noted in~\cite{Auger2} that the Auger phase measurements in adjacent energy intervals does not 
seem to be randomly distributed but rather suggests a smooth transition between a common phase of $\simeq 270^\circ$ 
(consistent with the right ascension of the galactic center) below $10^{18}$~eV and another phase (consistent with the right 
ascension of the galactic anti-center) above $5~10^{18}$~eV. This is shown in Figure~\ref{fig:harmonic2} (bottom panel).
This is potentially interesting, because with a real underlying anisotropy, a consistency of the phase measurements in ordered 
energy intervals is indeed expected to be revealed with a smaller number of events than needed to detect the amplitude with 
high statistical significance. It is noteworthy that some consistency can be observed with data coming largely from the Northern 
hemisphere, from the analysis of Yakutsk data, as shown in the top left panel of Figure~\ref{fig:harmonic2}, and from a recent 
analysis of TA data (top right panel). Such a smooth transition hence appears as an interesting feature to monitor with future data
from any UHECR observatory, to confirm whether this effect is genuine or not.

\section{Correlation with celestial objects}
\label{sec:3}

After almost 50 years since the first detection~\cite{Linsley:1963km} of cosmic rays of the order of 100~EeV, their origin and 
nature remain unknown. At the highest energies, the most probable sources of UHECRs are extragalactic~: jets of active galactic 
nuclei (AGN), radio lobes, gamma ray bursts and colliding galaxies, among others \cite{kotera}. The relevance of these objects 
as the UHECR sources may be tested by cross-correlation of the corresponding catalogues with the UHECR events. This section 
describes the correlation analyses performed by the Pierre Auger, HiRes and the Telescope Array collaborations.

\subsection{Correlation with the VCV catalogue}

\begin{figure}\centering
\includegraphics[width=0.48\linewidth]{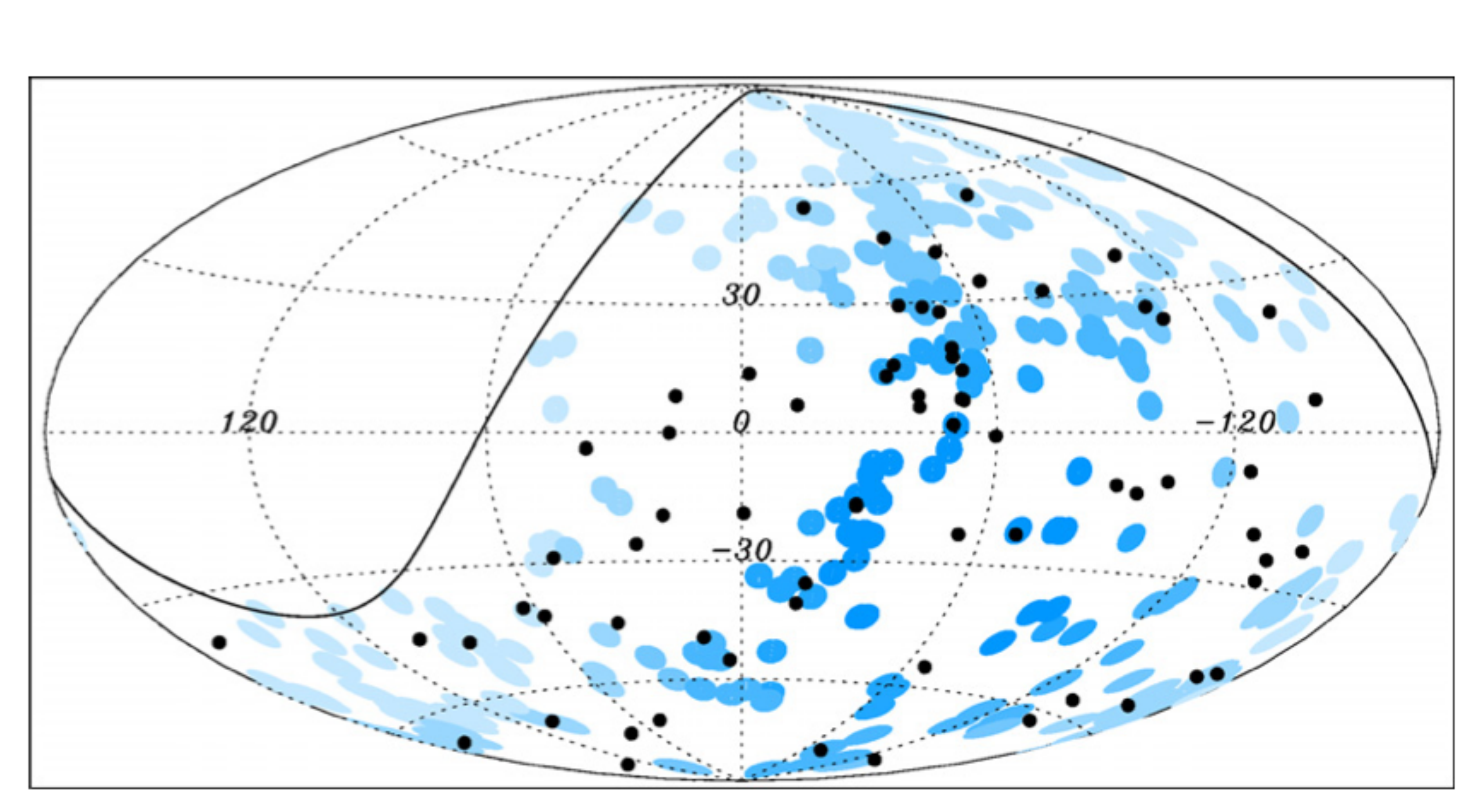}
\includegraphics[width=0.45\linewidth]{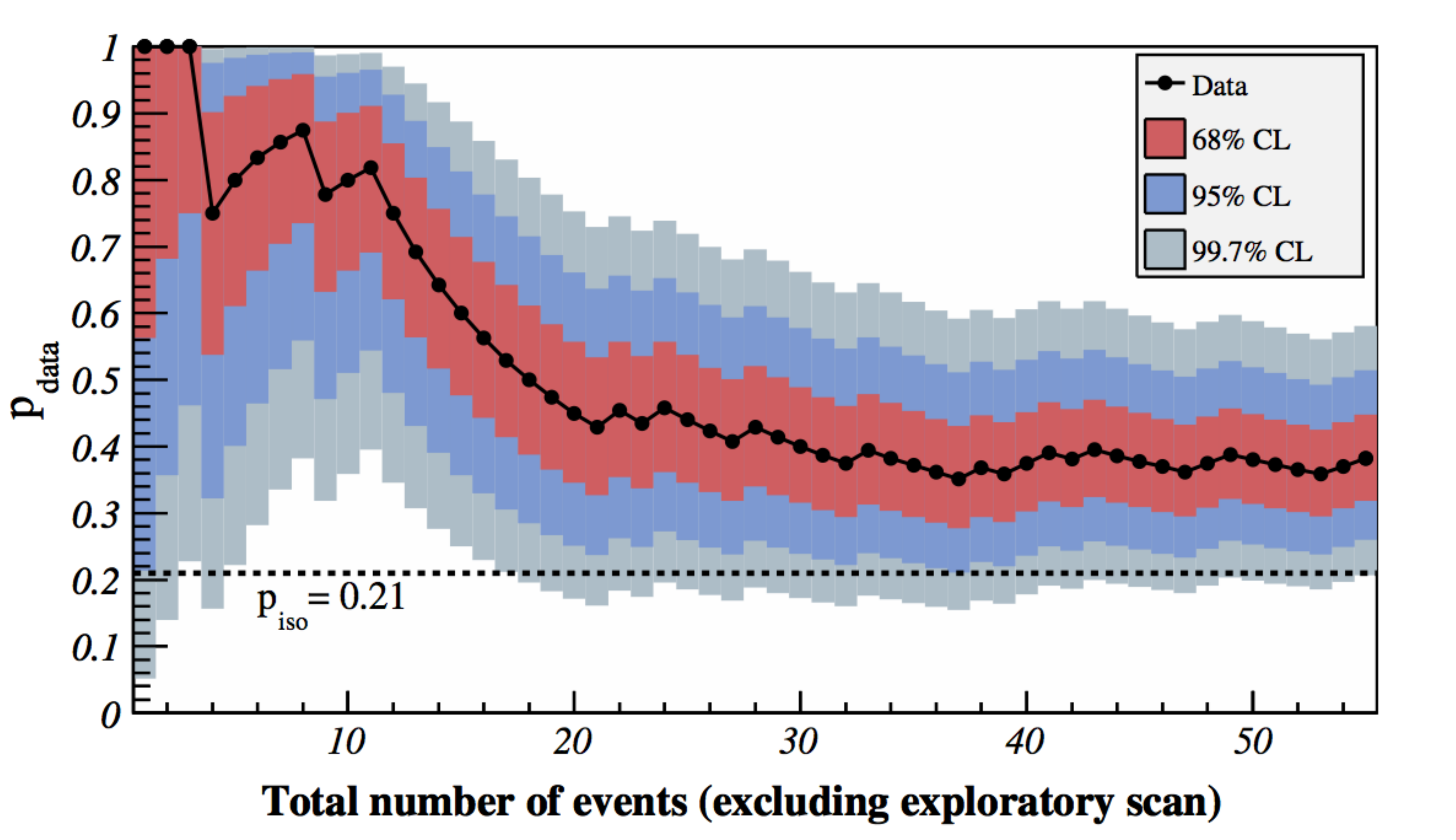}
\caption{Left~: The 69 arrival directions of CRs with energy $E>55$ EeV detected by the Pierre Auger Observatory up to 31 December 2009 are plotted as black dots in an Aitoff- Hammer projection of the sky in galactic coordinates. The solid line represents the border of the field of view of the Southern Observatory for zenith angles smaller than 60$^{\circ}$. Blue circles of radius 3.1$^{\circ}$ are centred at the positions of the 318 AGNs in the VCV catalog that lie within 75 Mpc and that are within the field of view of the Observatory. Darker blue indicates larger relative exposure. The exposure-weighted fraction of the sky covered by the blue circles is 21\%.   Right~:  Fraction of events correlating with AGNs as a function of the cumulative number of events, starting after the exploratory data. The expected correlating fraction for isotropic cosmic rays is shown by the dotted line.  From Ref. \cite{update}.}
\label{fig:sources1}
\end{figure}

\begin{figure}\centering
\includegraphics[width=0.48\linewidth]{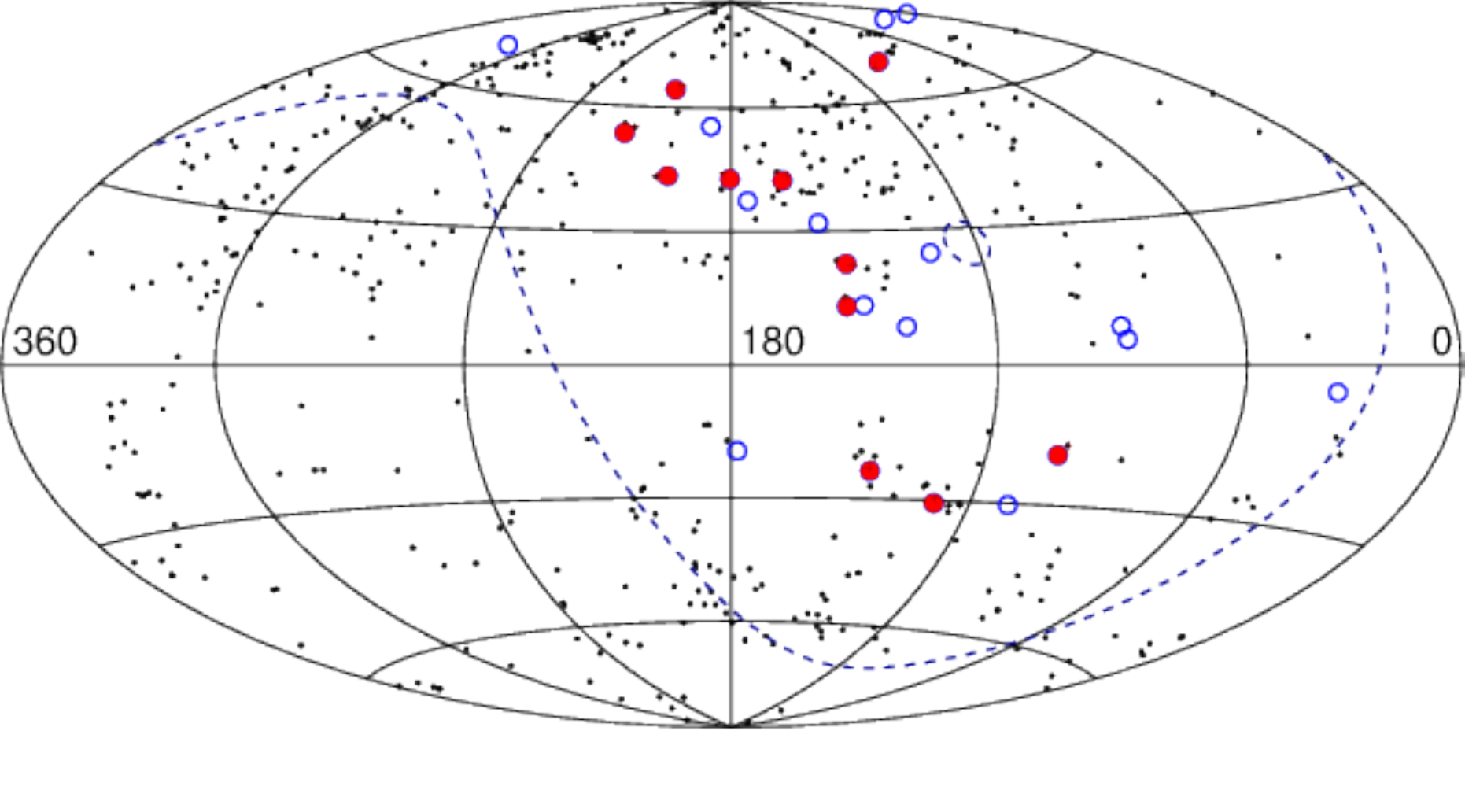}
\includegraphics[width=0.45\linewidth]{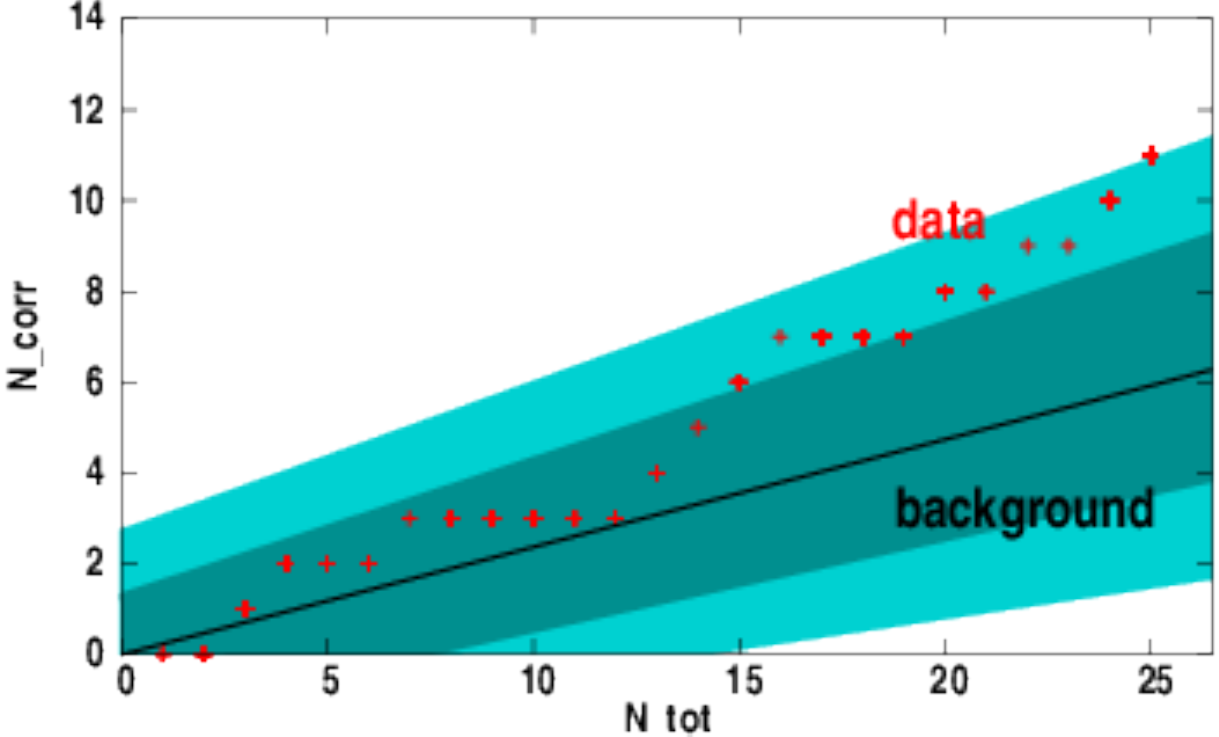}
\caption{Left~: Hammer projection of the TA cosmic ray events with $E > 57$ EeV and nearby AGNs in the Galactic coordinates. Correlating and non-correlating events are shown by filled red and empty blue circles, respectively. AGNs are represented by black dots. The dashed line shows the boundary of the TA exposure.   Right~:  The number of TA events with $E > 57$ EeV correlating with VCV AGNs as a function of the total number of events. The expectation for random coincidences is represented by the blue line together with the corresponding 68\%  and 95\% CL regions for isotropy. From Ref. \cite{TA}. }
\label{fig:sources2}
\end{figure}
 
The Pierre Auger Collaboration reported in~\cite{science,agn} a correlation of events with the AGNs in the Veron-Cetty \& Veron (VCV) catalogue~ \cite{VCV}. The first 14 events were used for an exploratory scan  that yielded the following search parameters~: energy threshold $E_{th} = 55$~EeV,  angular separation $\Psi \leq 3.1^{\circ}$, and redshift $z\leq 0.018$. Those parameters minimize the probability that the correlation with AGN could result from a background fluctuation if the flux were isotropic. The subsequent 13 events established a 99\% confidence level for rejecting the hypothesis of isotropic cosmic ray flux. The reported fraction of correlation events was $69^{+11}_{-13}$\%. 

Fig.~\ref{fig:sources1} (left panel) presents the updated high energy sky as measured at the Auger Observatory~\cite{update}, where the data with $E_{th} = 55$ EeV up to the end of 2009,  69 events in total, are shown by black dots. For this updated sample,  the amount of correlation decreased to $38^{+7}_{-6}$\%~\cite{update}. For this dataset, Fig.~\ref{fig:sources1} (right panel) shows the most likely value of the fraction of the correlated events plotted with black dots as a function of the total number of time-ordered events (the events used in the exploratory scan are excluded). The most updated estimate of the fraction of correlating cosmic rays is $33^{+5}_{-5}$\%  (28 events correlating from a total of 84 events)  and with 21\% expected under the isotropic hypothesis~\cite{kampertICRC}. The probability of this correlation to occur by chance with isotropic distribution of arrival directions is $P=0.006$. 

The HiRes Collaboration~\cite{HiresAGN} has searched for the correlation using search parameters prescribed by the Auger Collaboration in~\cite{science}. From the 13 events with energy above the search threshold, 2 events were found in correlations, while 3.2 were expected. Meanwhile, the most significant correlation found throughout a scan in $E_{th}$, $\Psi$, and $z$ occurred with a chance probability of 24\%.
                                                                                                                                                                                                                                                                                                                                                                                                                                                                                                                                                                                                                                                                                                                            
The Telescope Array Collaboration~\cite{TA} has also searched for this correlation. In this analysis, the 7 objects with zero redshift found in the VCV catalogue in the field of view of the TA were removed, so that 465 objects were left in the catalogue.  The TA exposure is peaked in the Northern hemisphere so the AGNs visible to TA are not the same as the ones visible to Auger, though there is some overlap. When the distribution of nearby AGNs is taken into account, and assuming equal AGN luminosities in UHECR, the correlating fraction would be $\sim$44\% following the updated correlation fraction from Auger.  In Fig. \ref{fig:sources2} (left panel), the sky map of TA events with energy $E>57$ EeV  and nearby AGNs from the VCV catalogue is shown in galactic coordinates.  The cosmic rays are shown in filled red (correlating events) and empty blue circles (non-correlating events). The AGNs are shown in black dots.  In Fig. \ref{fig:sources2} (right panel), the number of TA events correlating with AGNs is shown as a function of the total number of events with $E>57$ EeV ordered according to arrival time. The blue line stands for the expected number of random coincidences in case of a uniform distribution calculated via Monte Carlo simulation. The shaded region represent 68\% CL deviations from isotropy calculated by the maximum likelihood methof of Ref. \cite{gorbunov2006}. In the full TA SD data set, there are 11 correlating events out of 25, while the expected number of random coincidences is 5.9. The probability of this correlation to occur by chance  with isotropic distribution of arrival directions is $P=0.02$, since the probability of a single event to correlate is 24\% under the isotropic hypothesis.

The most recent values of the Auger correlating fraction and the TA correlating fraction are compatible. More data is necessary to show whether this correlation is statistically significant or not. 
 
\subsection{Correlation with Centaurus A}

Another possible scenario is that the anisotropy is dominated by cosmic rays originating from the vicinity of Centaurus A, the nearest active galaxy with an estimated distance of about 3.8 Mpc~\cite{cenA}. It was noted in~\cite{science,agn} that the arrival directions of two UHECR events correlate with the nucleus position of the Centaurus A, while several lie in the vicinity of its radio lobe extension.

In Ref.~\cite{update}, the Auger Collaboration points out that a circle of radius 18$^{\circ}$ centered on Cen A contains 13 of their highest energy events while only 3.2 would be expected from an isotropic flux. As this is an \textit{a posteriori} observation, no confidence level can be assigned to the excess in the Cen A region. Another issue is that Cen A is part of the nearby M83 Group of galaxies, which has behind the more distant Centaurus Supercluster with several clusters and groups of galaxies within the GZK sphere. So, the excess around Cen A cannot be cited as strong evidence that Cen A is a source of UHECRs. Let us note that the current overdensity observed around Cen A is partly responsible for the correlation with AGN found in the Auger data~\cite{Gorbunov2007}. It will be interesting to monitor this situation with additional data.

\subsection{Search for galactic neutron sources}

A source of ultra-high energy neutrons in the galaxy could be spotted since the mean path length for a neutron is $9.2 \times  E$  kpc before decaying, where $E$ is the energy of the neutron in EeV.  The Pierre Auger Collaboration has performed two analyses to constrain the neutron flux from Galactic sources in three energy bands~: [1-2]~EeV, [2-3]~EeV, and $E\geq 1$~EeV \cite{neutronICRC2011,neutrons}. In the first of them, a blind search for localized excesses in the CR flux over the exposed sky was performed initially.  The number of observed events was compared to the number of events expected from an isotropic background in top-hat counting regions matching the angular resolution of the detector. In the second, a stacking analysis was performed in the direction of bright Galactic gamma-ray sources~:  the ones detected by Fermi-Lat instrument in the 100~MeV - 100~GeV range and the ones detected by H.E.S.S. in the range of 100~GeV - 100~TeV. Both analyses didn't provide any evidence for significant excess and upper limits on the flux were derived for all directions within the Auger coverage. For directions along the Galactic plane for instance, the upper limits are below 0.024 km$^{-2}$yr$^{-1}$, 0.014 km$^{-2}$yr$^{-1}$
and 0.026 km$^{-2}$yr$^{-1}$ for the energy bins [1-2]~EeV, [2-3]~EeV and $E\geq1$~EeV, respectively.

Those upper limits call into question the existence of persistent sources of EeV protons in the Galaxy~\cite{neutrons}. They also place useful constraints on the UHE emissions from the known galactic sources of TeV gamma-rays. 

 \subsection{Correlations with BLLac}
 
Another possible scenario for sources of UHECRs are BL Lacertae objects, which are a sub-class of blazars, active galaxies with beamed emission from a relativistic jet which is aligned roughly toward our line of sight. Using a technique based on the angular correlation function, a correlation of the HiRes stereo data with a subset of BL Lacs on a scale of 0.8$^{\circ}$ consistent with its angular resolution was found in Ref.~\cite{gorbunov2004}.  This result was checked with a different analysis by the HiRes Collaboration~\cite{HiResBLL} and a correlation was found for events with $E>10$ EeV and the VCV catalogue. Nevertheless, data from the  Pierre Auger Collaboration did not confirm this correlation \cite{augerBLL}.

\section{Correlation with the large-scale structure of the Universe}

Given our current poor understanding of 
the composition of UHECRs, their sources and deflections on the way to
the Earth, one should look for signatures as robust as possible with
respect to these parameters. One of such signatures is the correlation
of UHECRs with the large-scale structure (LSS) of the Universe.

Both protons and nuclei are attenuated at highest energies in a
similar (although, not identical) way, so that at energies around
$10^{20}$~eV their propagation length is as short as a few tens of
Mpc. The observed cutoff in the spectrum is in agreement with this
picture. Thus, the sources of the highest-energy cosmic rays, whatever
they are, must be situated within this distance. Moreover, regardless
of their nature they must, at least at the largest scales, follow the
matter distribution.

At distances of a few tens of Mpc the matter distribution is
inhomogeneous forming the large-scale structure composed of galaxy
clusters, filaments and voids. Since the sources must be concentrated in
clusters, one expects the UHECR flux at highest energies to be
anisotropic, with  overdensities  (underdensities) in the directions
of nearby clusters (voids). 

The expected flux distribution is largely model-independent and
predictable provided the deflections of UHECRs in the magnetic fields
are known. Comparing the observed UHECR distribution with the one
predicted at different deflections one may, therefore, directly
measure or constrain the latter.

The UHECR deflections depend on both their charge composition and the
magnetic fields. Knowing one of these two factors would allow one to
access the other through correlations of UHECRs with LSS. For example,
given the existing bounds on the extragalactic fields
\cite{Kronberg:1993vk} and recent estimates of the Galactic one
\cite{Pshirkov:2011um} one can check that protons of energies of order
$6\times 10^{19}$~eV should not be deflected by more than $\sim 3-5^\circ$
except in the direction of the Galactic center. If stronger lower
bounds on deflections at corresponding energies will be derived, they
would suggest that the UHECR composition is predominantly heavy. In
other words, proton composition at the highest energies implies certain
correlation with the LSS. 

There are several methods to quantify correlations of UHECRs with the
LSS. The simplest one is a cross-correlation between the UHECR
events and the galaxy sample used to represent the matter
distribution. Such analysis has been performed by the Pierre Auger
collaboration \cite{Abreu:2010zzj}.  
Two catalogs were used to trace the matter distribution: 
2MRS catalog of galaxies \cite{2MRS} with $K_{\rm mag} < 11.25$ (this sample
contains $\sim 22\, 000$ galaxies within 200~Mpc), and 
the Swift-BAT hard X-ray catalog, containing 373 AGNs within 200~Mpc. 
Each CR arrival direction forms a pair with every object in the catalog. For
the cross-correlation estimator, the fractional excess 
(relative to the isotropic expectation) of pairs having angular separations 
smaller than any angle $\psi$ was used. This is given by 
$n_p(\psi)/n_p^{\rm iso} (\psi) -1$,
where $n_p(\psi)$ denotes the number of pairs with separation angle
less than $\psi$. 

\begin{figure}
\includegraphics[angle=-90,width=0.5\textwidth]{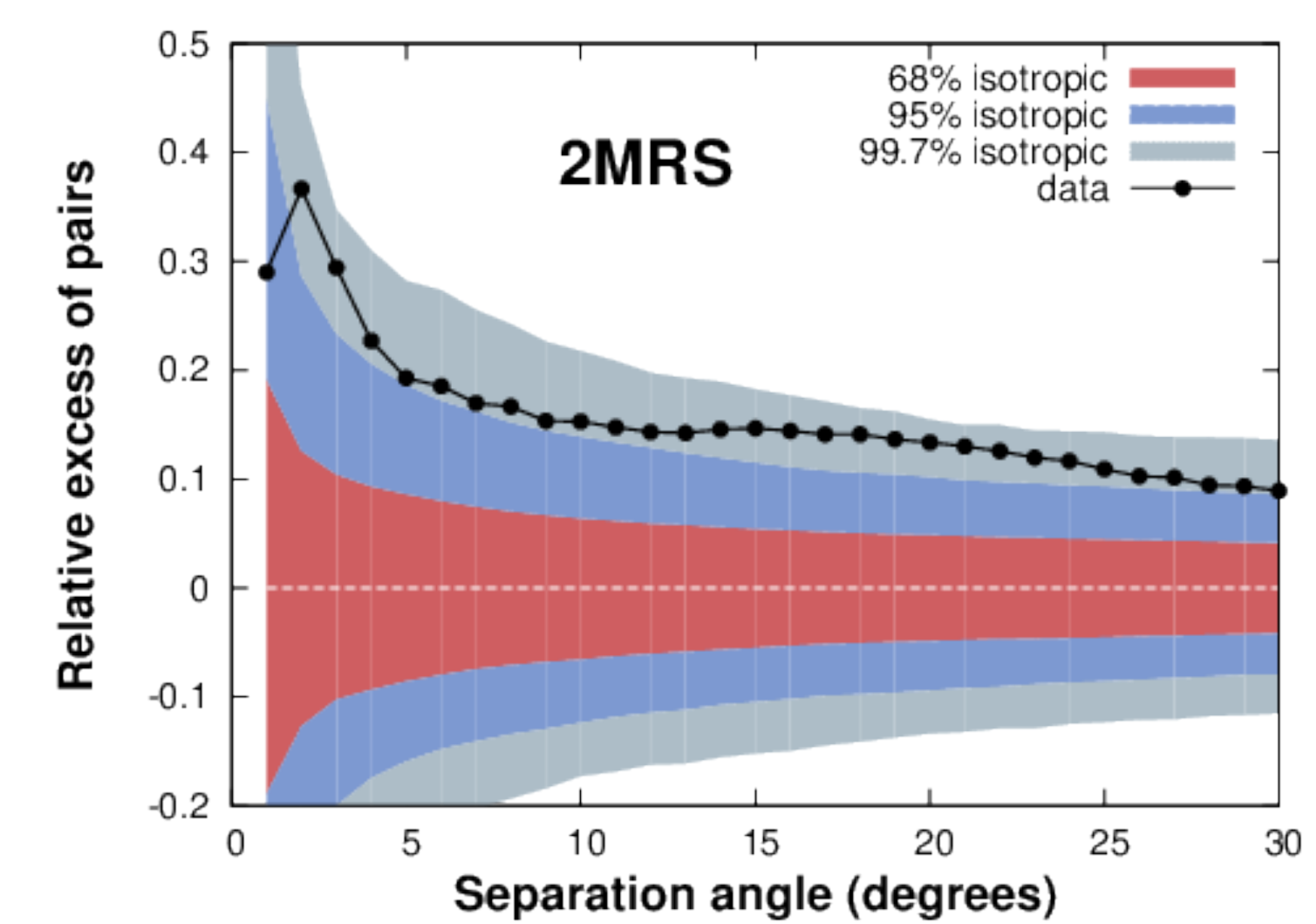}
\includegraphics[angle=-90,width=0.5\textwidth]{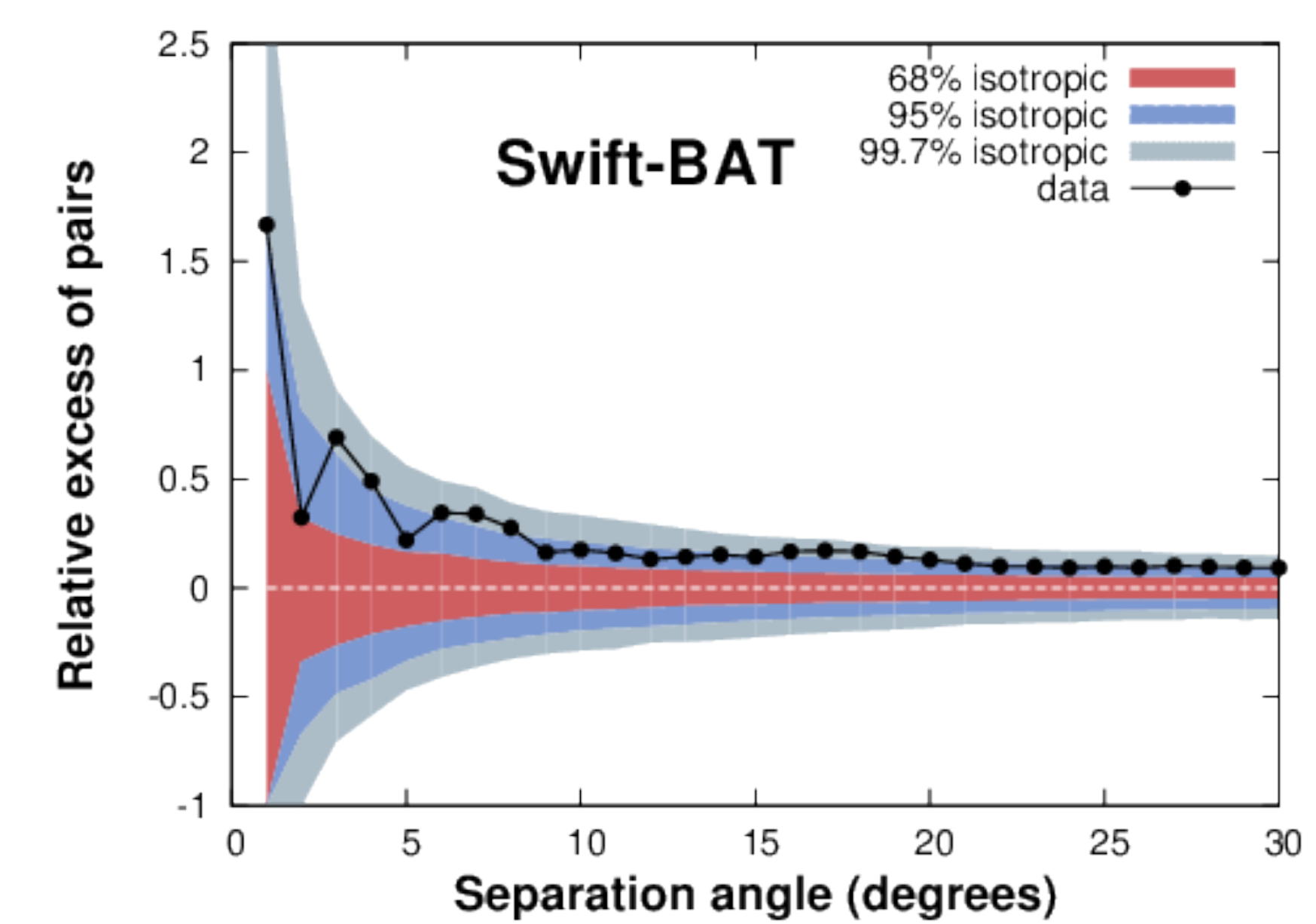}
\caption{Relative excess of pairs using data (black
dots) in the case of 2MRS galaxies (left) and Swift-BAT AGNs
(right). The bands contain the dispersion in 68\%, 95\%
and 99.7\% of simulated sets of the same number of events assuming 
isotropic cosmic rays. }
\label{fig:corr-2MRS}  
\end{figure}
Fig.~\ref{fig:corr-2MRS} shows the relative excess of pairs using data
(black dots) in the case of 2MRS galaxies (left) and Swift-BAT AGNs
(right). The bands in the plot contain the dispersion in 68\%, 95\%
and 99.7\% of simulated sets of the same number of events assuming
isotropic cosmic rays. The plots show the results using all the
arrival directions of CRs with $E\geq 55$~EeV collected between 1
January 2004 and 31 December 2009~: 69 CR events in the case of
correlation with Swift-BAT AGNs, and 57 CR events in the case of
correlation with galaxies in the 2MRS catalog (for which galactic
latitudes $|b| < 10^\circ$ were excluded).

The correlation in excess of isotropic expectations is observed in all
cases. However, the existence of cross-correlation does not imply that
the arrival directions are distributed in the sky in the same manner
as the objects under consideration, because the catalogs of astronomical
objects that were used here are flux-limited sets. 

A more accurate information about the source distribution can be obtained
if astronomical objects in the catalog are weighted according to the
flux of cosmic rays they produce in a given model. Such model is
defined by the catalog itself, the assumed intrinsic luminosities of objects
in UHECR and their spectrum, as well as the parameters controlling the
UHECR propagation. Given the model, the UHECR flux can be calculated
in the form of a smoothed flux density map and compared to the
observed distribution of events by means of an appropriate statistical
test. 

The analysis performed by the Pierre Auger Observatory uses the two
catalogs already discussed above: the 2MRS catalog of galaxies and
Swift-BAT hard X-ray catalog of AGN. The probability maps of arrival
directions of CRs expected in these models are constructed by
weighting the objects by their flux at the electromagnetic wavelength
relevant in the respective survey and by the attenuation factor
expected from the GZK effect. Contributions of individual objects are
smeared with the Gaussian distribution of a fixed angular width
$\sigma$ which represents typical (but unknown) deflections of UHECRs
in magnetic fields. An isotropically distributed fraction of events
$f_{\rm iso}$ is added in order to account for CR trajectories that
have been bent by wide angles due to large charges and/or encounters
with strong fields, or to the fraction of CRs arriving from beyond the GZK 
horizon. Thus, each of the two models contains two free
parameters: the smoothing angle $\sigma$ and the isotropic fraction
$f_{\rm iso}$.  More details on the models can be found in
Ref.~\cite{Abreu:2010zzj}. 

The model predictions are compared to the data by means of the binless
maximum likelihood method. The likelihood function is constructed as
the logarithm of the product of the probabilities to observe cosmic
ray in the directions $\hat {\bf n}_k$ of the actual events, 
\[
{\cal LL} = \sum _{k=1}^{N_{\rm data}} \ln F(\hat {\bf n}_k).
\]

\begin{figure}
\includegraphics[width=0.6\textwidth,height=0.28\textwidth]{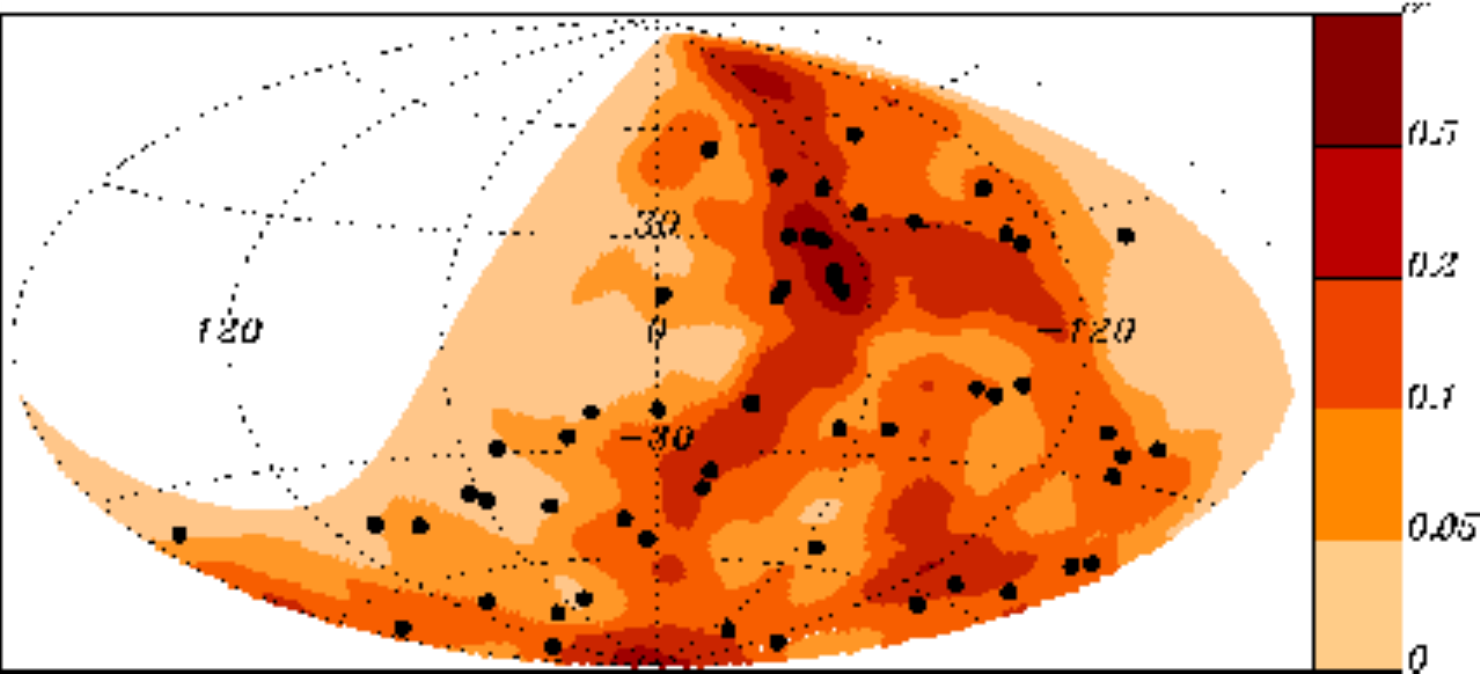}
\includegraphics[width=0.4\textwidth,height=0.28\textwidth]{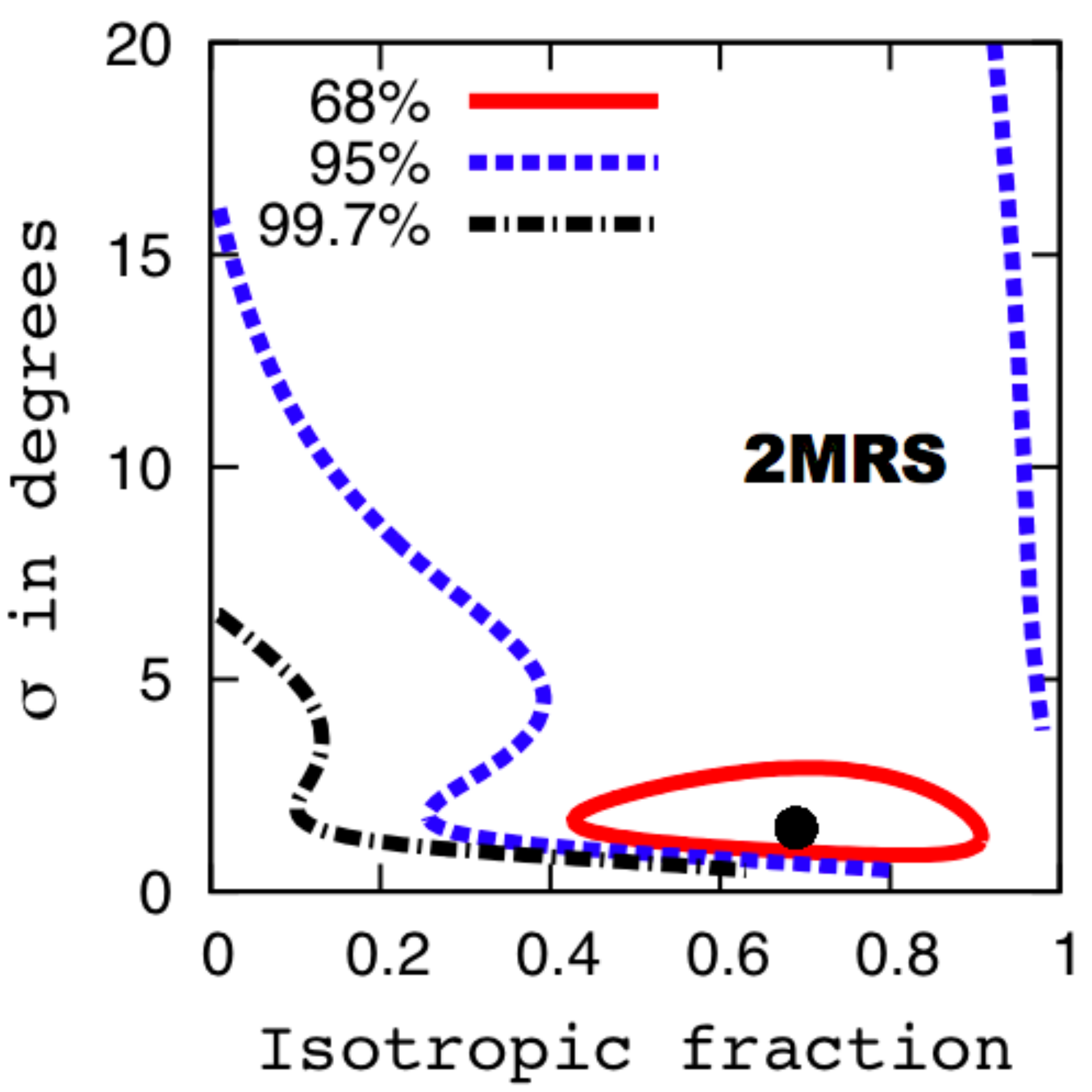}
\includegraphics[width=0.6\textwidth,height=0.28\textwidth]{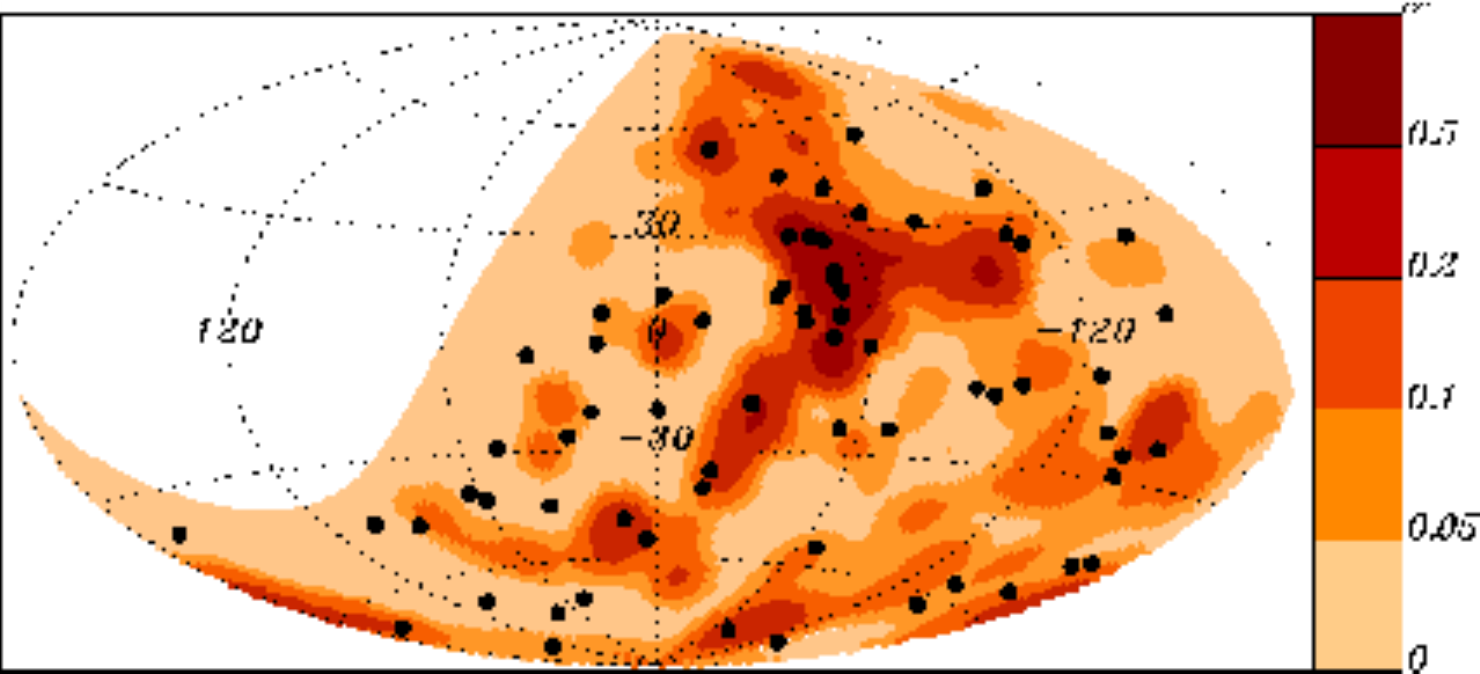}
\includegraphics[width=0.4\textwidth,height=0.28\textwidth]{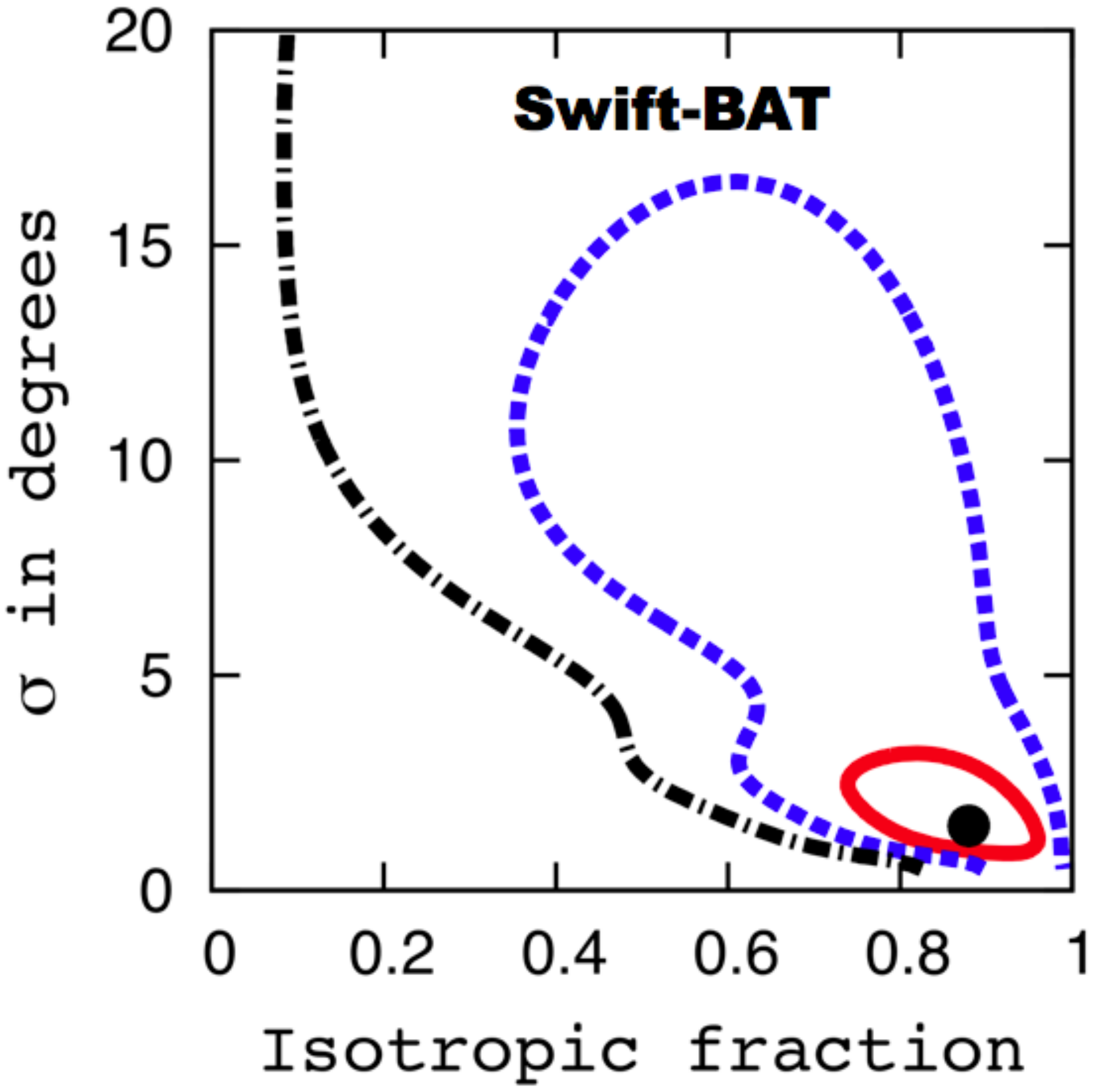}
\caption{{\it Left~:} Probability distribution of UHECR arrival directions in the case
  $E>55$~EeV and $\sigma=5^\circ$ for 2MRS galaxies (top) and
  Swift-BAT AGNs (bottom).  {\it Right~:} Confidence intervals
  for the parameters $(\sigma, f_{\rm iso})$ derived from the
  likelihood function for the two models: 2MRS galaxies (top) and
  Swift-BAT AGNs (bottom). Here we show only the plots that used data excluding the exploratory scan.   }
\label{fig:auger-skypl}  
\end{figure}
The results are presented in Fig.~\ref{fig:auger-skypl}. The left
panels show the smoothed flux distribution at $E>55$~EeV and $\sigma =
5^\circ$ for the 2MRS catalog and Swift-BAT catalog (top and bottom
rows, respectively).  The right panels show the contours of constant
likelihood function in the $(\sigma, f_{\rm iso})$ plane. 

The best-fit values of $(\sigma, f_{\rm iso})$ are (1.5$^\circ$, 0.69) for
2MRS and (1.5$^\circ$, 0.88) for Swift-BAT as indicated by a black
dot. Contours represent 68\%, 95\% and 99.7\% confidence intervals. (If the exploratory
scan period is included, the best-fit values of $(\sigma, f_{\rm iso})$ are (1.5$^\circ$, 0.64) for
2MRS and (7.8$^\circ$, 0.56) for Swift-BAT).
To test further the compatibility between the data and the models,
simulated sets with the same number of arrival directions as in the
data were generated, drawn either from the density map of the models
or isotropically. The distributions of the mean log-likelihood (${\cal
  LL} /N_{\rm data}$) thus obtained were compared to the value
obtained for the data.  The data are compatible with the models and
differ from average isotropic expectations. The fraction $f$ of
isotropic realizations that have a higher likelihood than the data is
about 0.02 for Swift-BAT AGNs and also for 2MRS galaxies. If the exploratory
scan period is included, the fraction is 
$2\times 10^{-4}$ in the case of the model based on Swift-BAT AGNs,
and $4\times 10^{-3}$ with the model based on 2MRS galaxies.  These
studies, however, are {\em a posteriori} and do not constitute further
quantitative evidence for anisotropy.

A similar (but not identical) analysis has been performed by the
Telescope Array collaboration making use of the surface detector data
collected in the period 2008.05.11-2011.09.1 (40 months). The model
that has been tested is the so-called matter-tracer model which
assumes that sources of UHECR are numerous and trace the matter
distribution in the Universe. The matter distribution was inferred
from the 2MASS Galaxy Redshift Catalog (XSCz) that is derived from the
2MASS Extended Source Catalog (XSC), with redshifts that have either
been spectroscopically measured (for most of the objects) or derived
from the 2MASS photometric measurements. The flux-limited subsample of
galaxies with apparent magnitude $m\leq 12.5$ was used in the UHECR
flux calculations. The objects closer than $5$~Mpc and further than
$250$~Mpc were removed from the sample in order to avoid breaking of
the statistical description. The contribution of the sources beyond
$250$~Mpc was replaced by a uniform flux normalized in such a way that
it provides the correct fraction of events as calculated in the
approximation of a uniform source distribution. The resulting catalog
contains 106\,218 galaxies, which is sufficient to accurately describe
the flux distribution at angular scales down to $\sim 2^\circ$. The
UHECR flux distribution is reconstructed from this flux-limited
catalog by the weighting method designed to compensate for the
selection effects (see Ref.\cite{AbuZayyad:2012hv} and references
therein).

All galaxies were assumed to have the same intrinsic luminosity in
UHECR and the same injection spectrum with the index equal to 2.4,
which is compatible with the UHECR spectrum observed by the HiRes and TA
\cite{AbuZayyad:2012ru} assuming proton composition and the source
evolution parameter $m=4$. The contributions of individual sources
were weighted according to their distances, and with full account of
the attenuation processes. Note that for solid angles at which the
statistical description holds (that is, such that the integrated flux
receives contributions from many sources for any direction), this
model is equivalent to the one used by the Auger collaboration based
on the 2MRS galaxy catalog.  The contributions of individual sources
were smeared with the Gaussian width $\theta$ which was considered a
free parameter and was varied in the range $2 - 20^\circ$. 

\begin{figure}
\resizebox{0.5\textwidth}{0.25\textwidth}{\includegraphics{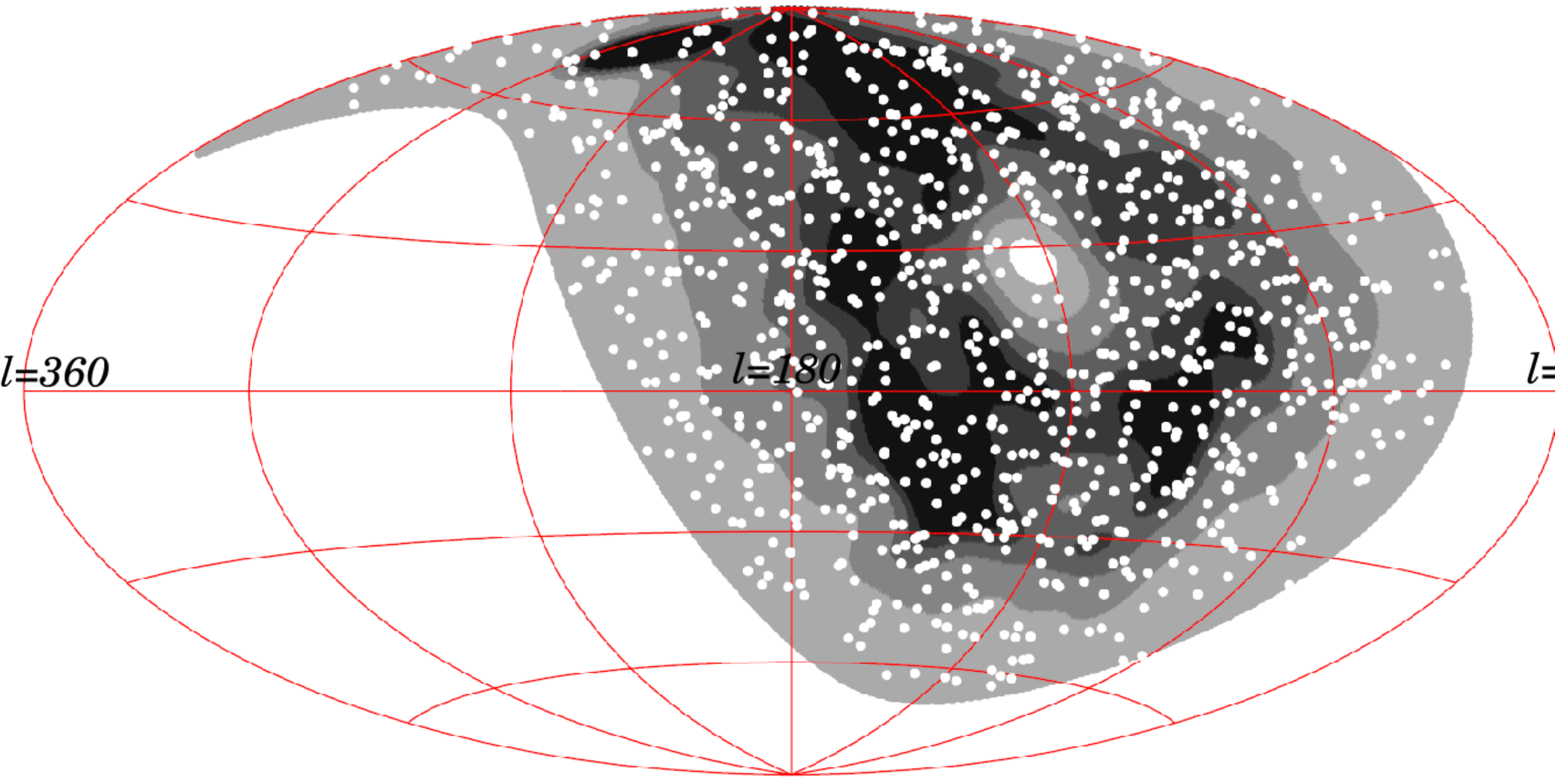} }
\resizebox{0.5\textwidth}{0.25\textwidth}{\includegraphics{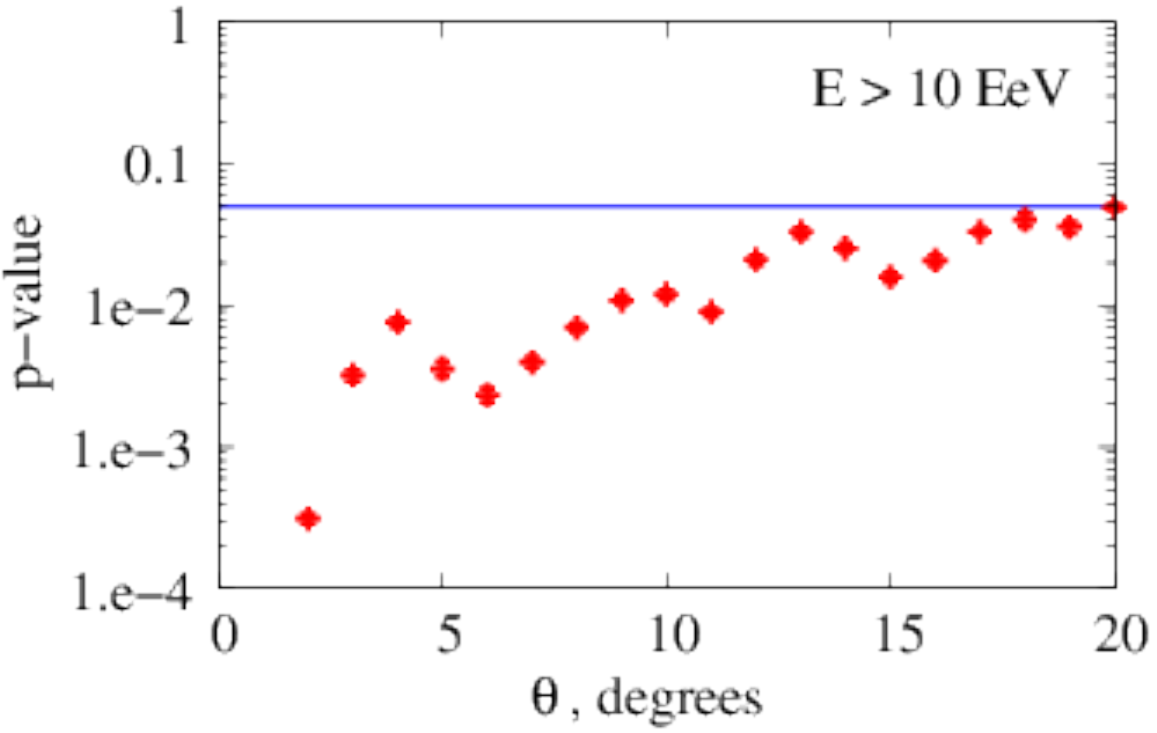} }
\resizebox{0.5\textwidth}{0.25\textwidth}{\includegraphics{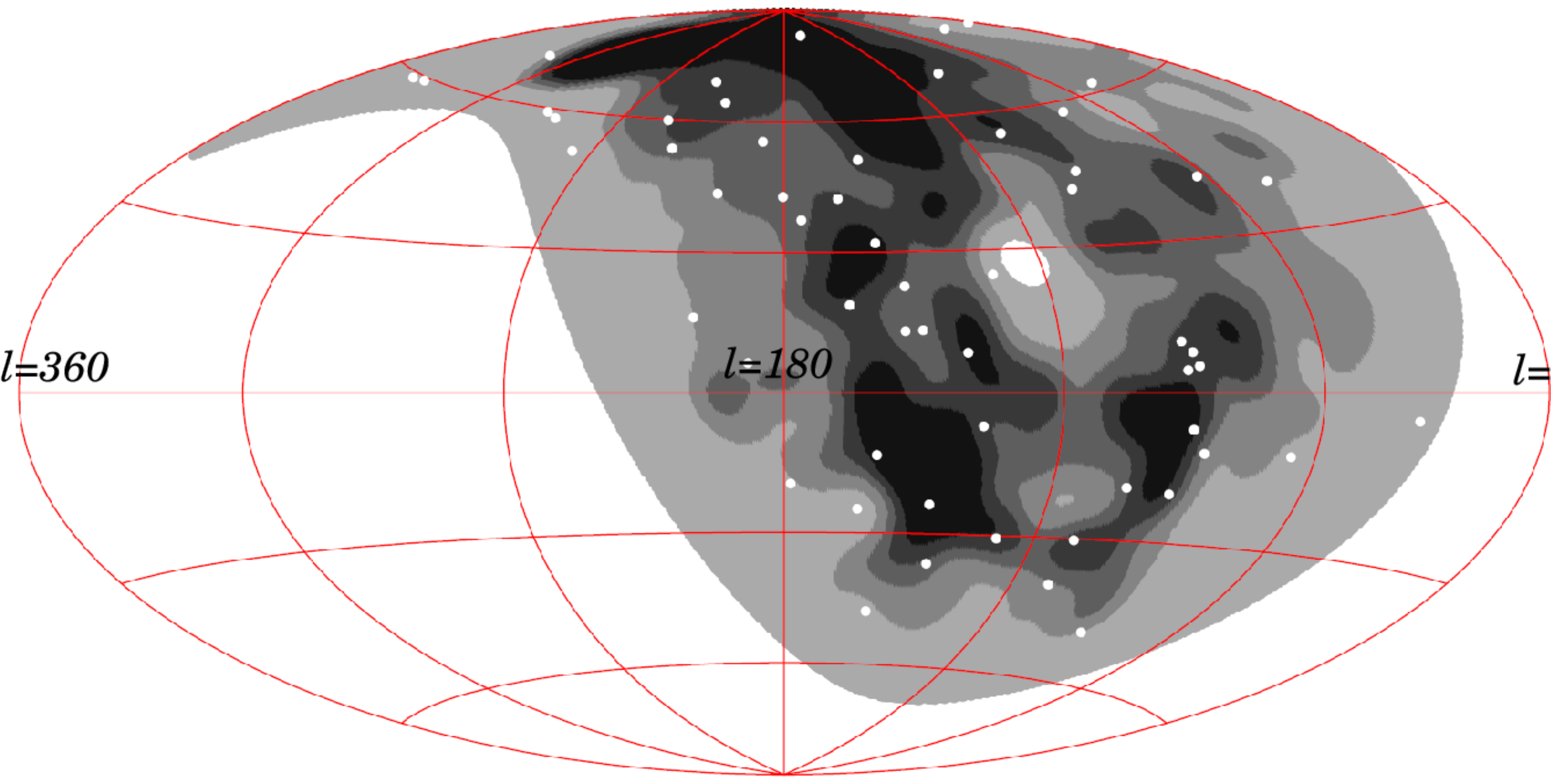} }
\resizebox{0.5\textwidth}{0.25\textwidth}{\includegraphics{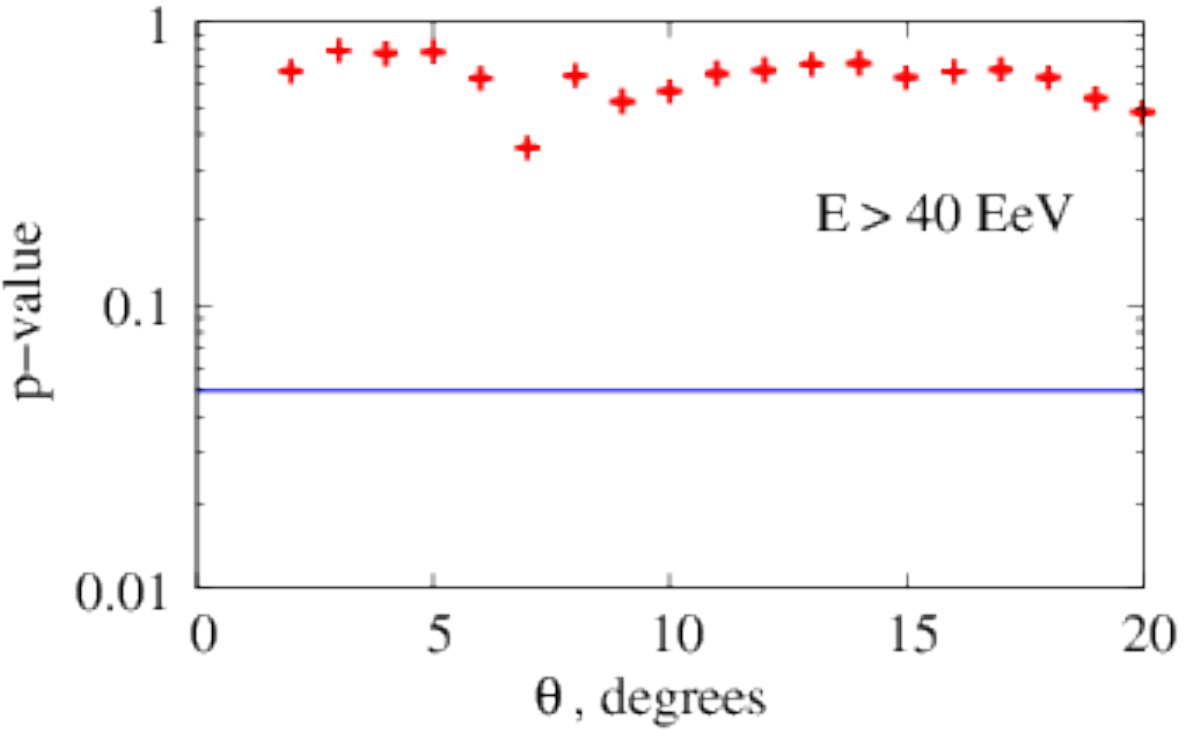} }
\resizebox{0.5\textwidth}{0.25\textwidth}{\includegraphics{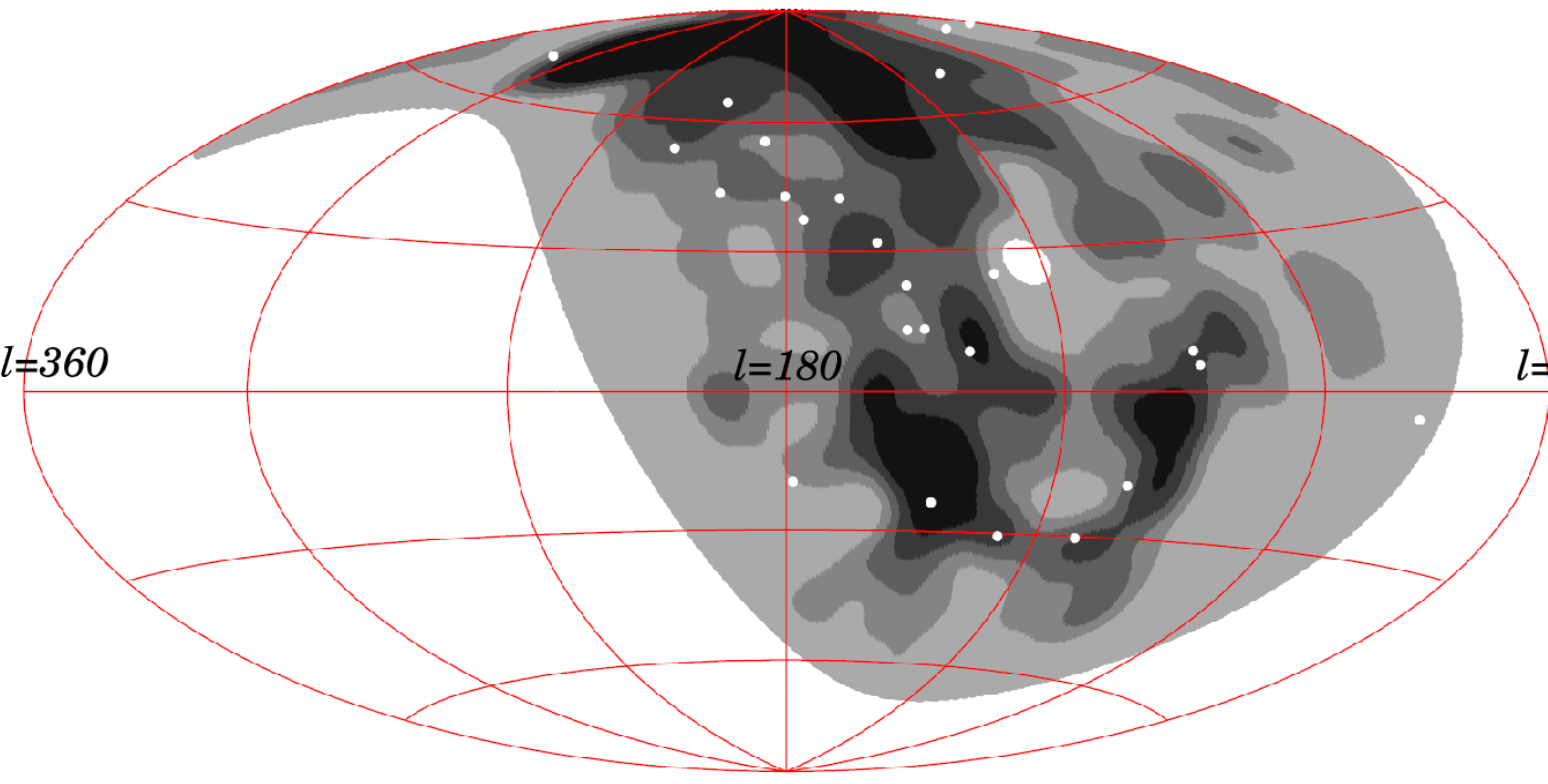} }
\resizebox{0.5\textwidth}{0.25\textwidth}{\includegraphics{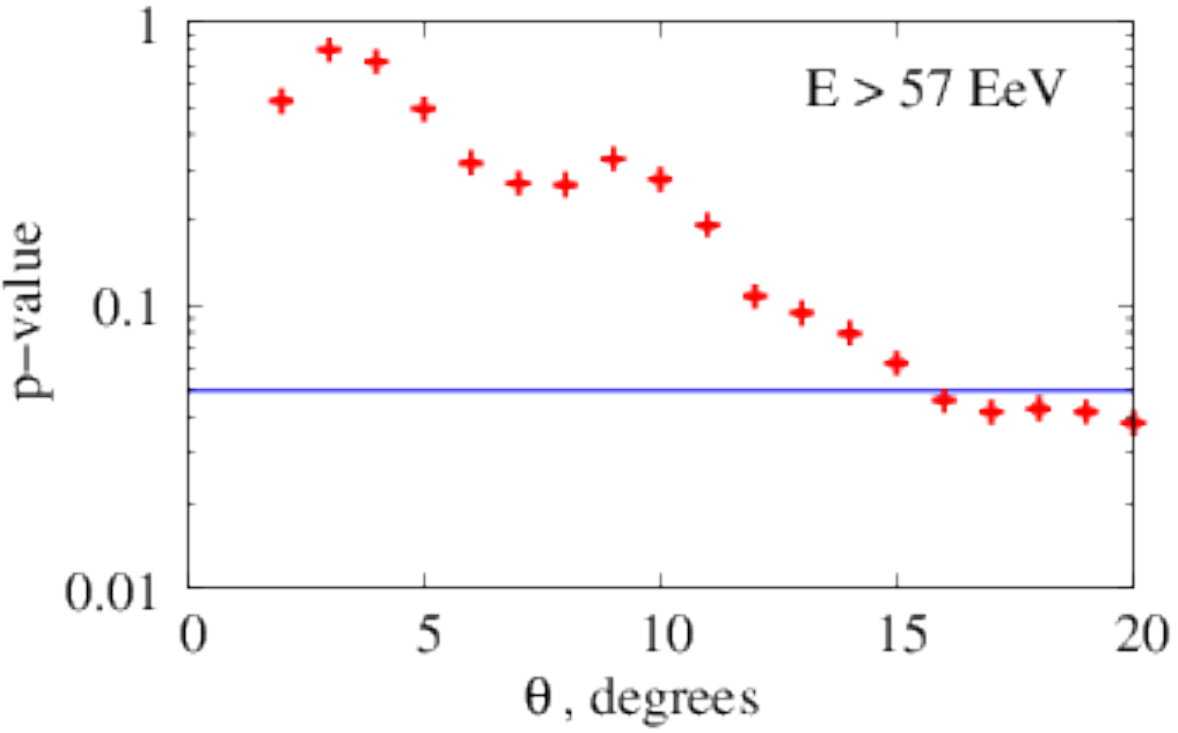} }
\caption{{\it Left column~:} The skymaps of the expected flux at energy thresholds of 10~EeV, 40~EeV, and 57~EeV (from top to bottom) in Galactic coordinates with the TA events superimposed (white dots) and a smearing angle of 6$^\circ$.{\it Right column:} The results of the statistical test for the compatibility between the data and the LSS hypothesis. Low $p-$value indicate incompatibility with the LSS model. The horizontal line shows a confidence level of 95\%. The three panels correspond to the same energy thresholds as in the left column.}
\label{fig:TA-skyplots}  
\end{figure}

Three {\em a priori} chosen energy thresholds were considered~:
$10$~EeV, $40$~EeV, and $57$~EeV.  For each energy threshold and each
value of the smearing angle $\theta$, the model flux map was compared
to the data by means of the flux-sampling test~\cite{Koers:2008ba}. 
The result of the test is the $p$-value which
characterizes the compatibility of the model and the data (low
$p$-values indicate incompatibility). These results are presented in
Fig.~\ref{fig:TA-skyplots}. 

As one can see from the plots, for
$E>40$~EeV and $E>57$~EeV the data are compatible with the
matter-tracer model at the 95\% C.L. At the energy threshold of
$E>10$~EeV the situation is somewhat different. The data are
incompatible with the matter-tracer model up to angles of order
$20^\circ$. This is not surprising in view of the expected large
deflections in magnetic fields at low energies.

Generating an isotropic flux map modulated with the TA exposure and
making use of the flux-sampling statistical test, one finds that for
all three energy thresholds $E>10$~EeV, $E>40$~EeV, and $E>57$~EeV the
TA data sets are compatible with isotropy at the 95\% C.L. The fact
that the high-energy event sets are compatible with both isotropy
and the matter-tracer model indicates insufficient statistics, as is
confirmed by the direct computation of the statistical power of the
flux-sampling test \cite{AbuZayyad:2012hv}. 

To summarize, although no firm conclusions about the anisotropy can be
derived from the analysis of correlations with the large scale
structure at present, interesting hints exist. Both the correlation
function and the likelihood method based on smoothed flux maps suggest
the departure from isotropy in the highest-energy Pierre Auger
data. The Telescope Array data, having smaller statistics, are
compatible with both isotropy and the LSS model, except at lowest
energies $E>10$~EeV where the data are not compatible with the LSS
model unless typical deflections at these energies exceed $\sim
20^\circ$, which is quite likely even for protons.

\section{Perspectives}

Though there are various hints of anisotropy in the arrival directions of UHECRs,
no clear signal has been established with certainty so far. As suggested by the analyses 
presented in this review, an order of magnitude increase in exposure is needed to fully
test these hints, or to find new kind of anisotropy and characterize its cause,
as well as to derive significantly stronger upper limits. Thus, a qualitative progress
seems to require a substantially improved UHECR detector that should have a 
number of features crucial for anisotropy searches. Most of the analyses presented
here are limited by statistics. The exposure of any future observatory
should thus be large enough to collect ten times the present world exposure in
five years after construction/deployment. 

There are sensible motivations to detect anisotropy in a broad energy range. As already
pointed out, establishing whether the intensity of extragalactic cosmic rays dominate the 
cosmic ray energy spectrum above some energy around $10^{18}$~eV would constitute 
an important step forward to provide further understanding on the origin of UHECRs. 
The transition from galactic to extragalactic cosmic rays should leave a signature in 
energy-dependent anisotropy at EeV energy or below. On the other hand, for extragalactic
cosmic rays above the GZK energy threshold, sources must be local and should follow the 
large scale structure of matter in the local universe. If there are protons, their arrival directions 
should correlate with that matter distribution on the sky. The order of magnitude increase of 
statistics is thus needed well below the ankle and also above the GZK-threshold~: the energy 
ranges from $10^{17.5}$~eV to $10^{20.5}$~eV (at least).

Anisotropy studies are intertwined with spectrum and composition
studies.  In particular, if the composition were known unambiguously,
it would be possible to design an observatory (and/or design an
analysis) that would make optimal cuts to increase the signal to
noise ratio, cut heavy nuclei for charged particle astronomy or neutron
astronomy, optimize cuts for filtering eventual gamma rays from the hadrons,
and so on and so forth. High discrimination power for composition on an 
event-by-event basis should be considered for a design of future
observatories.  

There would be many advantages to operate an observatory with full-sky
coverage. All possible point sources would be exposed. Obvious structure detection 
with minimal exposure distortion could be achieved. Besides, any anisotropy fingerprint 
is encoded in the spherical harmonic coefficients of the angular distribution of cosmic
rays. With full-sky coverage, the methods of spherical harmonics and multipole analysis 
could be applied without additional assumptions on the way the angular distribution
is behaving in the uncovered region of the sky. 

Finally, accurate energy assignments are essential for making precise energy cuts for 
trans-GZK anisotropy and also for large-scale anisotropy studies at EeV energies.
The relative energy resolution should not exceed 10\%, including statistical uncertainty 
of measurement, systematic uncertainty due to relative calibration, atmospheric profile,
temperature, pressure, clouds, aerosols, etc. As well, the absolute energy resolution should
not exceed 10\%, including fluorescence yield, absolute calibration, etc. The exposure 
uncertainty must be under control, never greater than the event count uncertainties.

\end{document}